\documentclass[12pt,journal,onecolumn,draftclsnofoot]{IEEEtran}
\newif\ifCLASSOPTIONromanappendices \CLASSOPTIONromanappendicestrue
\usepackage{tikz}
\usepackage[siunitx]{circuitikz}
\usepackage[utf8]{inputenc}
\usepackage{fontenc}
\usepackage{hyperref}
\usepackage{a0size}         
\usepackage{amssymb}
\usepackage{amsmath}
\usepackage{upgreek}
\usepackage{multicol}
\usepackage[english]{babel}     
\usepackage{epsfig}
\usepackage{bm}
\usepackage{oplotsymbl}
\usepackage{bbm}
\usepackage{enumitem}
\usepackage{amsfonts,color,amsthm,amsmath, lscape,graphics}
\usepackage{subfiles}
\usepackage[all]{hypcap}
\usepackage{comment}
\usepackage{algorithm} 
\usepackage{algpseudocode}

\usepackage[font=footnotesize]{caption}
\usepackage[font=footnotesize]{subcaption}
\usepackage{subcaption}
\usepackage{epstopdf}
\usepackage{algorithm} 
\usepackage{algpseudocode}
\usepackage{cite}

\definecolor{awesome}{rgb}{1.0, 0.13, 0.32}
\usepackage{fix2col}

\addto\captionsenglish{}

\newcommand*\xbar[1]{%
  \hbox{%
    \vbox{%
      \hrule height 0.5pt 
      \kern0.2ex
      \hbox{%
        \kern-0.1em
        \ensuremath{#1}%
        \kern-0.1em
      }%
    }%
  }%
} 

\usepackage{graphicx}  
\theoremstyle{plain}

\usepackage{multirow} 
\usepackage{relsize}  
\usepackage{textcomp, mathtools}
\usepackage{bm}   
\usepackage{pifont}
\usepackage{etoolbox}
\newtheorem{result}{Result}

\newcommand{\uvec}[1]{\boldsymbol{\hat{\textbf{#1}}}}
\makeatletter
\patchcmd{\@maketitle}
  {\addvspace{0.5\baselineskip}\egroup}
  {\addvspace{-1\baselineskip}\egroup}
  {}
  {}
\makeatother
\begin{document}
\title{Achievable Rate of Near-Field Communications Based on Physically Consistent Models}
\author{Mohamed Akrout, Volodymyr Shyianov, Faouzi~Bellili, \IEEEmembership{Member, IEEE},\\Amine Mezghani, \IEEEmembership{Member, IEEE}, Robert W. Heath, \IEEEmembership{Fellow, IEEE}
\thanks{The authors are with the Department of Electrical and Computer Engineering at the University of Manitoba, Winnipeg, MB, Canada (emails:\{akroutm,shyianov\}@myumanitoba.ca, \{Faouzi.Bellili,Amine.Mezghani\}@umanitoba.ca). R.~W.~Heath is at the North Carolina State University (email: rwheathjr@ncsu.edu). This work was supported by the Discovery Grants Program of the Natural Sciences and Engineering Research Council of Canada (NSERC) and the US National Science Foundation (NSF) Grant No. ECCS-1711702 and CNS-1731658.}}

\maketitle
\begin{abstract}
This paper introduces a novel information-theoretic approach for studying the effects of mutual coupling (MC), between the transmit and receive antennas, on the overall performance of single-input-single-output (SISO) near-field communications. By incorporating the finite antenna size constraint using Chu's theory and under the assumption of canonical-minimum scattering, we derive the MC between two radiating volumes of fixed sizes. Expressions for the self and mutual impedances are obtained by the use of the reciprocity theorem. Based on a circuit-theoretic two-port model for SISO radio communication systems, we establish the achievable rate for a given pair of transmit and receive antenna sizes, thereby providing an upper bound on the system performance under physical size constraints. Through the lens of these findings, we shed new light on the influence of MC on the information-theoretic limits of near-field communications using compact antennas.
\end{abstract}
\section{Introduction}\label{Section 1}
\subsection{Background and motivation}

Near-field (NF) communication has diverse applications from chip-to-chip \cite{kim2015miniaturized} and board-to-board \cite{gulati2019experimental} communications to Internet-of-Things (IoT) applications like  medical implantable devices \cite{won2021wireless,shin2017flexible}. Yet, the increasingly complex circuitry of interconnect and compact antenna design of these systems is a major data rate bottleneck in improving their overall performance. Moreover, the close proximity between the transmit and receive antennas in NF communications leads to strong interaction and the potential for a significant mutual coupling (MC) between their induced reactive electromagnetic fields. This alters the electromagnetic behavior of both antennas thereby widening the gap between the theoretically predicted performance when the MC is ignored and the practically achieved performance of near-filed communication systems \cite{bird2015fundamentals,bird2021mutual}. In this context, the strong NF interactions together with the significant MC effects must be meticulously studied to ultimately characterize the physical limitations of near-field wireless systems. The latter cannot be gauged by accounting for the MC effects only. In fact, the antenna size must also be considered as an integral part of the actual performance analysis of wireless NF communication systems. This is because the footprint of today’s wireless technology is dominated by the antenna size, which cannot be miniaturized beyond the Chu limit \cite{chu1948physical} without a degradation in the operational bandwidth.

\noindent The analysis and design of communication systems have historically evolved around the basic percept of separating the physical and mathematical abstractions of communication theory\footnote{Particularly the celebrated Shannon capacity formula for band-limited additive white Gaussian noise (AWGN) channels.}\cite{shannon1948mathematical} based on different scientific principles. Several recent attempts to reconnect these areas include \textit{wave theory of information} \cite{franceschetti2017wave}, \textit{electromagnetic information theory} \cite{gruber2008new, migliore2008electromagnetics}, and \textit{circuit theory  for communication} \cite{ivrlavc2010toward}.
Multiple studies in wave radiation and propagation confirmed that circuit and electromagnetic field theories are essential for the analysis and design of near-field communication systems \cite{wallace2004mutual, phang2018near, svantesson2001mutual}. Most of these studies, however, are limited to narrow-band communications, and very few of them \cite{gustafsson2004spectral,shyianov2021achievable} have so far considered the antenna size as a physical constraint in their respective designs to characterize the achievable rate of communication systems. Moreover, their analysis ignored the mutual coupling effect due to the far-field assumption. When the operational frequency band is low, the near-field range is wide enough to enable communication through inductive coupling \cite{azad2012link}, as used in tap-and-pay credit card applications. However, in the radiative near-field and far-field regions, only communication through the energy radiation process is possible. Taken together, this calls for a principled approach applicable in all frequency bands and communication regions while systematically incorporating the MC in a physically-consistent model \cite{ivrlavc2010toward} starting at the Maxwell’s equations level~\cite{jackson1999classical}.

\noindent

Physically-consistent models \cite{ivrlavc2010toward} refer to a class of models connecting the governing physics of antennas with the mathematics of communication systems using circuit theory. Such models are becoming popular because they provide a careful modeling of the antenna device within the RF chain (e.g., matching network, LNA). This is key to understanding the realistic physical limits on the achievable rate and develop optimization schemes to optimize it for wideband/compact antenna design \cite{saab2019capacity,saab2019capacitybased}.\\
Physically-consistent models come into play as an effective tool to describe the underlying circuits of the near-field communication models. This is because they can incorporate the transmit/receive antenna circuits (along with their associated self/mutual impedances) into the wireless communication channel in contrast to standard wireless communication channels where only the propagation effects (e.g., fading, shadowing, scattering) are considered. Moreover, in any physically-consistent model, the noise cannot be regarded from a statistical viewpoint only, e.g., treating it as an additive Gaussian-distributed random variable. There is a need for a precise description of the noise correlation at the receiver depending on the noise sources at both the transmiter and the receiver along with the self/mutual impedances of the antennas.

\subsection{Contributions}


In this paper, we combine Chu's \cite{chu1948physical} and canonical minimum scattering (CMS) antennas \cite{kahn1965minimum} to develop a physically-consistent description of the wireless communication channel and the noise correlation. This allows us to account for antenna size limitations using Chu’s theory \cite{chu1948physical} and MC effects together independently of the antenna type. By operating at the lowest transverse magnetic (TM) radiation mode only, Chu's CMS antennas have the broadest bandwidth compared to all the Chu’s antennas excited by higher-order radiation modes, thereby allowing to benefit from their simplicity while being consistent with the physical constraints on the antenna size. Therefore, the Chu approach enables us to decouple the analysis from a specific antenna design yet capture the salient features of the broadest possible antenna that fits within a prescribed sphere of radius $a$.

We characterize the MC by deriving the mutual impedances between two Chu's CMS antennas assumed confined in a spherical volume of radius $a_{\text{T}}$ and $a_{\text{R}}$. While the NF mutual impedances are commonly derived using the induced EMF method \cite{orfanidis2002electromagnetic} for CMS antennas, applying it to the Chu's CMS antennas is not possible because their electromagnetic fields are only known outside their encompassing spheres. To overcome this limitation, we first establish an equivalence between the radiated fields of a Chu's CMS antenna and a Hertz dipole using the equivalence theorem. After applying the induced EMF method to two Hertz dipoles, we use the obtained Chu-Hertz equivalence to deduce the NF mutual impedances between two Chu's CMS antennas. We also derive their FF mutual impedance based on the Friis's transmission equation.

Based on the obtained equivalent circuit models, we derive the mutual information between the input and output signals of the system under the MC and the antenna size constraint for reactive and radiative NF and FF communications. We find that the resulting signal-to-noise ratio (SNR) depends
strongly on the antenna size and that the performance degradation is more pronounced when the antenna size becomes a small fraction of the wavelength. For the colinear and parallel orientations of a SISO communication system, we show that the achievable rate of the colinear orientation drops below the one of the parallel configuration in the radiative NF region. Finally, we illustrate that the optimal power allocation provides a gain of up to 50\% in terms of achievable rate in the NF region and over 200\% in the FF region.

\subsection{Organization of the paper and notations}
We structure the rest of this paper as follows. In Section~\ref{sec:preliminaries}, we introduce the relevant preliminaries of circuit theory for SISO communication and Chu's theory \cite{chu1948physical} for size-constrained antenna characterization. In Section~\ref{sec:siso-system}, we present the circuit-theoretic SISO communication model and derive its input-output relationship for a given mutual impedance matrix. We then specialize the channel of the circuit-theoretic model to both near- and far-field SISO communications. In Section \ref{sec:NF-FF}, we compute the near- and far-field mutual impedance matrices using the equivalence theorem between CMS antennas and Hertz dipoles. Then, in Section~\ref{sec:achiveable-rate-optimization}, we derive the achievable rate with uniform as well as optimal power allocation strategies. Finally, our simulation results are presented in Section~\ref{sec:results} for both colinear and parallel antenna orientations, from which we draw out some concluding remarks in the near- and far-field regions.
\newline 
\noindent We also mention the common notation used in this paper. Given any complex number $z$,  $\Re[z]$, $\Im[z]$, and $\xbar{z}$ return its real part, imaginary part, and complex conjugate.  The statistical expectation and variance are denoted as $\mathbb{E}[\cdot]$ and $\textrm{Var}[\cdot]$. We also use $j$ to denote the imaginary unit (i.e., $j^{2}=-1$) and the symbol $\times$ to refer the cross product between two vectors. For a signal $u(t)$, we denote its Fourier transform by $U(f)$. Throughout the paper, the derivation of the transfer functions in the Fourier domain is assuming finite-energy signals, and we only consider by analytical continuation general random stochastic processes once we invoke the power spectral density (PSD).  Moreover, $c$ denotes the speed of light in vacuum (i.e., $c \approx 3\times10^8$), $\lambda$ is the wavelength, and $k_b = 1.38 \times 10^{-23}\, \mathrm{m}^{2}\, \mathrm{kg} \,\mathrm{s}^{-2}\, \mathrm{K}^{-1}$ is the Boltzmann constant. $\mu = 1.25\times 10^{-6} \,\mathrm{m}\, \mathrm{kg} \,\mathrm{s}^{-2}\, \mathrm{A}^{-2}$ and $\epsilon = 8.85\times 10^{-12} \,\mathrm{m}^{-3}\, \mathrm{kg}^{-1} \,\mathrm{s}^{4}\, \mathrm{A}^{2}$ are the permeability and permittivity of vacuum. Finally, $k=\omega\sqrt{\epsilon \, \mu}=\frac{2\pi}{\lambda}$ and $\eta=\sqrt{\frac{\mu}{\epsilon}}$ are the wave number and the wave impedance of a plane wave in free space.

\section{Preliminaries}\label{sec:preliminaries}
In this section, we present the two-port circuit model associated with the NF and FF SISO communication channel. We then recall the achievable rate of a SISO wireless communication channel used as the criterion to evaluate the performance of the overall performance of a NF SISO communication system in Section~\ref{sec:results}. We also present the equivalent circuit of the CMS antenna based on Chu's theory. In Section \ref{sec:siso-system}, we will incorporate this antenna circuit to describe the circuit model for NF communications.

\subsection{Circuit theory for communication}
The physically consistent study of random signals that are transmitted through wireless channels uses a circuit-theoretic approach (see \cite{shyianov2021achievable} and \cite{ivrlavc2010toward} for more details).
From circuit theory, transmitted/received signals are either voltages or currents that flow through the ports of the transmit/receive antennas. Finding the relationship between port variables at the transmitter(s) and receiver(s) is key to modelling both near- and far-field communication channels in a physically consistent way.
\vspace{-.6cm}
 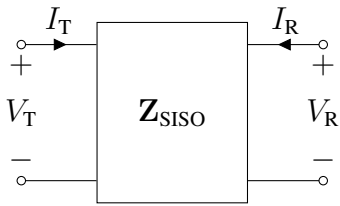
\begin{figure}[h!]
    \centering
    \begin{circuitikz}[american voltages, american currents, scale=1, transform shape]
\draw (0,0)
node[draw,minimum width=2cm,minimum height=2.4cm] (load) {$\mathbf{Z}_{\textrm{SISO}}$}
  ($(load.west)!0.75!(load.north west)$) coordinate (la)
  ($(load.west)!0.75!(load.south west)$) coordinate (lb)
  ($(load.east)!0.75!(load.north east)$) coordinate (lc)
  ($(load.east)!0.75!(load.south east)$) coordinate (ld)
  
  ($(lb) + (-1,0)$) to[short,o-] (lb)
  ($(la) + (-1,0)$) to[short,i>^=$I_\text{T}$,o-] (la)
  ($(ld) + (1,0)$) to[short,o-] (ld)
  ($(lc) + (1,0)$) to[short,i>_=$I_\text{R}$,o-] (lc)
  
  ($(la) + (-1,0)$) to [open,v=$V_\text{T}$] ($(lb) + (-1,0)$)
  ($(lc) + (1,0)$) to [open,v^=$V_\text{R}$] ($(ld) + (1,0)$);
  \vspace{1cm}
\end{circuitikz}
    \caption{Equivalent circuit-theoretic model for SISO communication channels.}
    \label{fig:channel-circuit-models}
    \vspace{-0.25cm}
\end{figure}

\noindent A SISO communication channel can be viewed as a \textit{two-port network} as depicted in Fig. \ref{fig:channel-circuit-models} connecting the port currents, $(I_\text{T}(f),I_\text{R}(f))$, and the port voltages, $(V_\text{T}(f),V_\text{R}(f))$, through an impedance matrix $\mathbf{Z}_{\textrm{SISO}}$ as 
\begin{equation}\label{eq:siso-V=ZI}
    \Bigg[\begin{array}{l}
V_\text{T}(f) \\
V_\text{R}(f)
\end{array}\Bigg]~=~\underbrace{\Bigg[\begin{array}{cc}
\text{Z}_{\text{T}}(f) & \text{Z}_{\text{TR}}(f)\\
\text{Z}_{\text{RT}}(f) & \text{Z}_{\text{R}}(f)
\end{array}\Bigg]}_{\mathbf{Z}_{\text{SISO}}(f)}\,\Bigg[\begin{array}{l}
I_\text{T}(f) \\
I_\text{R}(f)
\end{array}\Bigg].
\end{equation}

\noindent The diagonal entries, $\text{Z}_{\text{T}}$ and  $\text{Z}_{\text{R}}$, represent the self-impedances of transmit and receive antennas. They correspond to input impedances of the antennas when these are hypothetically isolated (i.e., when each antenna is considered alone). The off-diagonal entries, $\text{Z}_{\text{RT}}$ and  $\text{Z}_{\text{TR}}$, represent the mutual transmit-receive and receive-transmit impedances, respectively, between transmit and receive antennas. For Chu's CMS antennas, the impedance matrix is calculated analytically in Section~\ref{sec:NF-FF}, thereby leading to a compact two-port matrix description of SISO communication systems along with their easy analysis.

\subsection{Achievable rate of SISO wireless communication channels}
The achievable rate of a continuous-time additive Gaussian noise channel with a certain transmit band-limited power spectral density, $P_{\rm t}(f)$, is given by:
\begin{equation}\label{eq:AWGN_Capacity}
    C = \int_{0}^{\infty}{\log_2\left(1+\frac{P_{\rm t}(f)\,|H(f)|^2}{N(f)}\right)\textrm{d}f}\,\,[{\textrm{bits/s}}],
\end{equation}
 where $H(f)$ and $N(f)$ are the Fourier transform of the channel $h(t)$ and the PSD of the noise. In (\ref{eq:AWGN_Capacity}), it is assumed that $N(f)$ is either white or integrable, i.e., $\int_{-\infty}^{+\infty}N(f)\, \textrm{d}f< \infty$. Note that by letting $|H(f)|^2 = 1$ and $N(f) = N_0$ in (\ref{eq:AWGN_Capacity}), one recovers the well-known capacity of the AWGN channel. This is a standard result found in most information theory textbooks, e.g. \cite{gallager1968information}. It is adopted hereafter in the context of circuit-theoretic modelling of NF and FF SISO communications as the key criterion to evaluate the system performance using the Chu's CMS antennas. This is as opposed to the optimisation of the S-parameters which are the conventional antenna’s figures of merit that illustrate its circuit behavior.
 
 \noindent The achievable rate in (\ref{eq:AWGN_Capacity}) accounts for the MC effects and the antenna size when both $H(f)$ and $N(f)$ are derived using the equivalent circuit-theoretic models of communication systems. In this case, $H(f)$ and $N(f)$ are functions not only of the propagation conditions but also of the self/mutual impedances of transmit/receive antennas as will be shown in Section~\ref{sec:achiveable-rate-optimization}. In this case, the PSD of the noise $N(f)$ is still integrable but not white, thereby justifying the valid use of (\ref{eq:AWGN_Capacity}) throughout the paper.
 

\subsection{The antenna size constraint from Chu's theory}
To incorporate the constraint on the antenna size appropriately, we resort to Chu’s seminal work in \cite{chu1948physical}. There, Chu lays the foundations for the equivalent circuit models representing the superposition of $\text{TM}_{\text{n}}$ radiation modes of a linearly polarized antenna that is embedded inside a spherical volume of a given radius $a$. After deriving the electromagnetic (EM) fields outside the sphere enclosing the antenna, Chu described the equivalent circuit for each radiation mode $\text{TM}_{\text{n}}$ and uniquely determined its associated voltage $V_\textrm{n}(f)$ and current $I_\textrm{n}(f)$. We review in Appendix~\ref{appendix:Chu-radiated-EM-fields} the radiated EM fields of Chu's antennas at the details needed for a comprehensive exposure of their circuit-equivalent representations.

\begin{figure}[h!]
\centering
\begin{circuitikz}[american voltages, american currents, scale=0.8, every node/.style={transform shape}]
\draw (0,0) node[anchor=east]{}
 to[short, o-*] (3,0);
 \draw (3,2) to[L, label=\mbox{\small{$L=\frac{a \,R}{c}$}}, *-*] (3,0);
 \draw (3,0) -- (5,0);
 \draw (5,2) to[/tikz/circuitikz/bipoles/length=30pt, R, l=\mbox{\small{$R$}}, -] (5,0);
 \draw (3,2) -- (5,2);
 \draw (0,2) node[anchor=east]{}
  to[C, i>_=\mbox{\small{$I_1(f)$}}, label=\mbox{\small{$C=\frac{a}{cR}$}}, o-*] (3,2);
  \draw[-latex] (1,0.5) -- node[above=0.05mm] {$Z_\text{Chu}(f)$} (2, 0.5);
  \draw (0,2) to [open,v=\small{$V_1(f)$}] (0,0);
\end{circuitikz}
\caption{Equivalent circuit for the TM$_1$ mode of Chu’s electric antennas.}
\label{fig:tm1}
\end{figure}
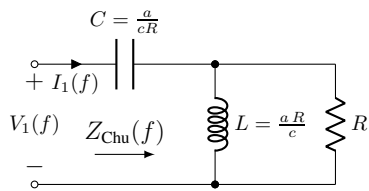

\noindent In wideband communications, antennas are expected to have the lowest Q-factor which can only be achieved when they are operating at their lowest radiation mode. Indeed, the higher the radiation mode, the larger the stored (i.e., non-radiated) energy in the sphere enclosing the antenna. It is therefore enough to consider the lowest radiation mode only\footnote{Combining the lowest magnetic and electric modes, $\text{TM}_{\text{1}}$ and $\text{TE}_{\text{1}}$, a.k.a., magneto-electric antennas \cite{hansen2011small}, can improve the Chu's limit and thus achieve a better bandwidth. For simplicity consideration, we only consider the electric antenna only.}, i.e., $n=1$. The corresponding port voltage and current, $V_1$ and $I_1$, of the antenna are expressed as \cite{chu1948physical}

\begin{subequations}\label{appendix-eq:V-I-TM1}
    \begin{align}
    V_{1}(f)&~=~\sqrt{\frac{8 \pi \eta}{3}} \,\frac{\sqrt{R}\,A_{1}}{k} \, \bigg(1 + \frac{1}{jka} - \frac{1}{j(ka)^2}\bigg)\,e^{-jka}~[\text{V}], \\
I_{1}(f)&~=~-\sqrt{\frac{8 \pi\eta}{3}}\,\frac{A_{1}}{\sqrt{R}\,k} \, \bigg(1 + \frac{1}{jka}\bigg) \,e^{-jka}~[\text{A}],
    \end{align}
\end{subequations}

\noindent where $k=\frac{2\pi\,f}{c}$ and $A_1$ is the complex coefficient of the $\textrm{TM}_1$ mode and $R$ is the resistance of the antenna. In this case, the equivalent  circuit for the TM$_1$ wave or the so-called 
``\textit{Chu's electric antenna}'' 
is illustrated in Fig.~\ref{fig:tm1}.

\noindent Using basic circuit analysis, one can establish the expression of the associated input impedance
\begin{equation}\label{eq:Z1-chu}
    Z_\text{Chu}(f) ~=~ \frac{V_1(f)}{I_1(f)}~=~ \underbrace{\frac{1}{j\,2\pi f\,\frac{a}{c\,R}}}_{Z_C(f)} + \underbrace{\frac{1}{\frac{1}{j\,2 \pi f\,\frac{a\,R}{c}}+\frac{1}{R}}}_{Z_{\rm L}(f) \parallel\, Z_R(f)} = \frac{c^2R + j\,2\pi\, f\,c\,a\,R - (2\pi \,f\,a)^2\,R}{j\, 2\pi \,f \,c\,a - (2 \pi \,f\,a)^2}\, [\Omega].
\end{equation}
We will rely in this paper on the circuit model of the Chu's CMS antenna depicted in Fig.~\ref{fig:tm1} to obtain the maximum achievable rate for any transmitter and receiver of fixed sizes $a_{\text{T}}$ and $a_{\text{R}}$.


\section{SISO communication system model}\label{sec:siso-system}
In this section, we describe the circuit-theoretic model for SISO communications. First, we explain the relationships between the noise sources in both NF and FF scenarios. We then derive the generic input-output relationship of the channel model as function of the self and mutual impedances of the transmit/receive antennas independently of the choice of the antenna. Next, we incorporate transmit/receive Chu's CMS antennas to describe the circuit model associated with wideband SISO NF and FF communications. This highlights the flexibility of circuit models \cite{ivrlavc2010toward} to provide a systematic approach to analyze both NF and FF communications, unlike the communication models for NF communications through inductive coupling \cite{azad2012link}.

\subsection{A circuit-theoretic SISO communication model}\label{subsec:siso-model}
\begin{figure}[h!]
\centering
\begin{circuitikz}[american voltages, american currents, scale=0.75, every node/.style={transform shape}]
\draw (-4.5,2) to[/tikz/circuitikz/bipoles/length=33pt, V, l_=$V_{\textrm{T}}$] (-4.5,0);
\draw (-4.5,2) to[/tikz/circuitikz/bipoles/length=20pt,R,l=$R$] (-3,2);
\draw (-4.5,0) to[short,-o] (-3,0);
\draw  (-1.5, 2) to[/tikz/circuitikz/bipoles/length=30pt,V, label=\mbox{}, l_=$\widetilde{V}_{\textrm{N,T}}$, o-o] (-3, 2);
\draw (-3,0) to[short, o-o] (-1.5,0);
\draw (-3, 2) to [open,v=$V_1^\prime$] (-3,0);
\draw (-1.5,0) to[short] (-0.5,0);
\draw (-1.5,2) to [short,i>^=$I_1$, o-] (-0.5,2);
\draw (0.5,1) node[draw,minimum width=2cm,minimum height=2.4cm] (load) {$\mathbf{Z}_{\textrm{SISO}}$};
\draw (-1.5, 2) to [open,v=$V_1$] (-1.5,0);
\draw (2.5, 2) to [open,v=$V_2$] (2.5,0)
(2.5,2) to [short,i>_=$I_2$] (1.5,2)
(2.5,0) -- (1.5,0)
(4, 2) to [open,v=$V_2^\prime$] (4,0);
\draw  (4,2) to[/tikz/circuitikz/bipoles/length=30pt,V, label=\mbox{}, l_=$\widetilde{V}_{\textrm{N,R}}$, -o] (2.5,2);
\draw (4,2) to[short, o-] (4.5,2);
\draw (4.5,2)  to[/tikz/circuitikz/bipoles/length=20pt,R, l^=\mbox{{$R_{{\textrm{in}}}$}}, i>^=\mbox{{$I_{R_{{\textrm{in}}}}$}}, -*] (4.5,0);
\draw (2.5,0) to[short,o-*] (4.5,0);
\draw (4,0) to[short,o-*] (6,0);
\draw  (6, 2) to[/tikz/circuitikz/bipoles/length=30pt,cV, label=\mbox{}] (6, 0);
\draw  (7.5,2) to[/tikz/circuitikz/bipoles/length=30pt,V, label=\mbox{},l_=$\widetilde{V}_{\textrm{N,LNA}}$, o-] (6,2);
\draw (6,0) to[short,-o] (7.5,0);
\draw (7.5,2) to [open,v={{$V_R$} }] (7.5,0);
\node[] at (6.8,0.7) {\small{$V_{\text{LNA}}$}};
\draw [dashed] (-1.5, -0.1) to (-1.5, -1.5);
\draw [dashed] (2.5, -0.1) to (2.5, -1.5);
\draw [dashed] (4, -0.1) to (4, -1.5);
\draw [dashed] (-3, -0.1) to (-3, -1.5);
\node[] at (-4.4,-0.5) {$\small{\textrm{Transmit signal}}$};
\node[] at (-4.4,-0.9) {$\small{\textrm{generator}}$};
\node[] at (0.5,-0.8) {$\small{\text{the wireless channel}}$};
\node[] at (3.25,-0.4) {$\small{\textrm{Receiver}}$};
\node[] at (3.25,-0.8) {$\small{\textrm{extrinsic}}$};
\node[] at (3.25,-1.2) {$\small{\textrm{noise}}$};
\node[] at (6,-0.75) {$\small{\textrm{Receiver LNA}}$};
\node[] at (-2.25,-0.4) {$\small{\textrm{Transmit}}$};
\node[] at (-2.25,-0.8) {$\small{\textrm{extrinsic}}$};
\node[] at (-2.25,-1.2) {$\small{\textrm{noise}}$};
\end{circuitikz}
\caption{SISO communication model including the signal generator, the transmit and receive extrinsic noises, the communication channel with antenna mutual coupling, and the LNA model with the associated intrinsic noise. The output voltage  $V_\text{R}$ is connected to the load impedance of a receive device. The frequency argument was dropped to lighten the notation in the figure.}
\label{fig:siso-system-model-nomn}
\vspace{-0.3cm}
\end{figure}
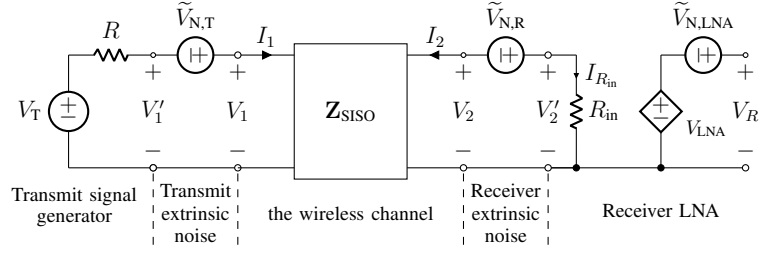

We consider the circuit-based model, depicted in Fig. \ref{fig:siso-system-model-nomn}, to study a SISO communication system where the transmit and receive antennas are at either NF or FF separation distances. The associated wireless channel is represented by the frequency-domain impedance matrix $\mathbf{Z}_{\text{SISO}}$ given in (\ref{eq:siso-V=ZI}), which model both the NF and FF communication channels $\mathbf{Z}_\text{SISO}^{\textrm{NF}}$ and $\mathbf{Z}_\text{SISO}^{\textrm{FF}}$, depending on whether the MC effects are taken into account or not. For more realistic scenarios, we hereafter consider a noisy communication channel of the noiseless model in (\ref{eq:siso-V=ZI}).

\subsubsection{Incorporating noise into the noiseless communication  channel  model}
Accounting for the noise sources at the transmit and receive antennas boils down to injecting their background noise at both the input and output ports. This can be done when the two-port network is only composed of passive components which have the same absolute temperature $T$ as the surrounding environment \cite{ivrlavc2010toward}. This situation is called \textit{thermal equilibrium noise} and corresponds to the case where the two-port network noise originates solely from the thermal agitation of electrons as they flow inside its all-passive components.
To this end, we introduce two voltage sources $\widetilde{V}_{\textrm{N,T}}(f)$ and $\widetilde{V}_{\textrm{N,R}}(f)$ at the transmitter and receiver ports as depicted in Fig.~\ref{fig:siso-system-model-nomn}. There, the terminals of the non-ideal generator voltage $V_{\text{T}}(f)$ (i.e., with internal resistance $R$) are connected to the transmitting antenna through a noisy input port with the current-voltage pair $(V_1(f),I_1(f))$. Likewise, the noisy receive antenna terminals are connected to the outside world through the output port with current-voltage pair $(V_2(f),I_2(f))$.

\noindent When the mutual coupling is ignored (e.g., the FF scenario), the noise voltages $\widetilde{V}_{\textrm{N,T}}(f)$  and $\widetilde{V}_{\textrm{N,R}}(f)$ are uncorrelated from each other, i.e.:
\begin{equation}\label{eq:cross-correlation-zero}
\mathbb{E}\left[\widetilde{V}_{\rm{N,T}}(f)~ \xbar{\widetilde{V}_{\rm{N,R}}(f)}\,\right] = 0. 
\end{equation}
\noindent Besides, the PSDs $S_{\widetilde{V}_{\rm{N,T}}}(f)$ and $S_{\widetilde{V}_{\rm{N,R}}}(f)$ of the noise voltage signals $\widetilde{V}_{\rm{N,T}}(f)$ and  $\widetilde{V}_{\rm{N,R}}(f)$ with finite energy, are given by \cite{ivrlavc2010toward,nyquist1928thermal}
\begin{subequations}\label{eq:auto-correlation-NF-FF}
\begin{align}
    S_{\widetilde{V}_{\rm{N,T}}}(f) &= 4\,k_\text{b}\,T\,\Re{[\text{Z}_{\text{T}}(f)]},\\
    S_{\widetilde{V}_{\rm{N,R}}}(f) &= 4\,k_\text{b}\,T\,\Re{[\text{Z}_{\text{R}}(f)]}.
\end{align}
\end{subequations}

\noindent In NF communications, the noise voltages $\widetilde{V}_{\textrm{N,T}}(f)$  and $\widetilde{V}_{\textrm{N,R}}(f)$ are correlated due to the MC induced by the non-zero mutual impedances between the transmit and receive antennas. The cross-correlations $S_{\widetilde{V}_{\rm{N,T}}, \widetilde{V}_{\rm{N,R}}}(f)$ and $S_{\widetilde{V}_{\rm{N,R}}, \widetilde{V}_{\rm{N,T}}}(f)$ between $\widetilde{V}_{\textrm{N,T}}(f)$  and $\widetilde{V}_{\textrm{N,R}}(f)$ are given by \cite{ivrlavc2010toward,nyquist1928thermal}\vspace{-0.3cm}
\begin{subequations}\label{eq:cross-correlation-NF}
\begin{align}
    S_{\widetilde{V}_{\rm{N,T}}, \widetilde{V}_{\rm{N,R}}}(f) &= 4\,k_\text{b}\,T\,\Re{[\text{Z}_{\text{TR}}(f)]},\\
    S_{\widetilde{V}_{\rm{N,R}}, \widetilde{V}_{\rm{N,T}}}(f) &= 4\,k_\text{b}\,T\,\Re{[\text{Z}_{\text{RT}}(f)]}.
\end{align}
\end{subequations}
\vspace{-0.5cm}

Together, the impedance matrix and the noise covariance matrix provide the circuit-theoretic description of the NF and FF communication channels modelled in Fig.~\ref{fig:siso-system-model-nomn}. This model encapsulates the transmit and receive antenna responses together with the NF and FF propagation aspects, hence the convenience of circuit-theoretic models for communication systems analysis \cite{ivrlavc2010toward}.
\subsubsection{The receive LNA model} The LNA is modeled as a noisy frequency-flat device with gain $\beta$, with the input-output voltage relationship
\begin{equation}\label{eq:LNA-voltage}
    V_{\rm{LNA}}(f) = \beta \,V_{2}^\prime(f)\,\,[{\textrm{V}}].
\end{equation}
For an amplifier with an input impedance $R_{\textrm{in}}$ and a noise figure $N_\textrm{f}$, we compute the PSD of the intrinsic noise voltage, $\widetilde{V}_{\textrm{N,LNA}}(f)$, generated inside the LNA as $4\,k_\text{b}\,T\,R_{\textrm{in}}\,(N_\text{f} - 1)$ \cite[chapter 10]{pozar2011microwave}. 
Moreover, the amplifier noise voltage, $\widetilde{V}_{\rm{N},\rm{LNA}}(f)$, is uncorrelated with the transmit noise voltage, $\widetilde{V}_{\rm{N,T}}(f)$, and the receive noise voltage, $\widetilde{V}_{\rm{N,R}}(f)$.
\subsubsection{The input-output relationship of the channel model} Adding the noise to both the input and output ports of the SISO communication channel model transforms its linear input-output relationship (\ref{eq:siso-V=ZI}) into an affine one. By applying Kirchhoff's voltage law (KVL) in Fig.~\ref{fig:siso-system-model-nomn}, we obtain an affine noisy two-port communication channel model:
\begin{equation}\label{eq:affine-siso-V=ZI}
    \Bigg[\begin{array}{l}
V_{1}^{\prime}(f) \\
V_{2}^{\prime}(f)
\end{array}\Bigg]~=~\mathbf{Z}_{\text{SISO}}\Bigg[\begin{array}{l}
I_{1}(f) \\
I_{2}(f)
\end{array}\Bigg] + \Bigg[\begin{array}{l}
\widetilde{V}_{\rm{N,T}}(f) \\
\widetilde{V}_{\rm{N,R}}(f)
\end{array}\Bigg].
\end{equation}

\noindent Moreover, using the definitions of $V_1^{\prime}(f)$ and $V_2^{\prime}(f)$ from (\ref{eq:affine-siso-V=ZI}), and $V_{\rm{LNA}}(f)$ from (\ref{eq:LNA-voltage}) and basic circuit analysis, the relationship between the output voltage $V_{\rm{R}}(f)$ and the input voltage $V_{\text{T}}(f)$ is obtained as follows:
\begin{equation}\label{eq:siso-input-output-relationship}
    V_{\textrm{R}}(f) = \widetilde{V}_{\text{N,LNA}}(f) + \beta\,R_{\text{in}}\,\frac{Z_{\text{TR}}(f)\,\big(V_{\text{T}}(f) + \widetilde{V}_{\textrm{N,T}}(f)\big) + (R+Z_{\text{T}})\,\widetilde{V}_{\textrm{N,R}}(f)}{\big(R_{\text{in}}+Z_{\text{R}}(f)\big)\,\big(R+Z_{\text{T}}(f)\big) - Z_{\text{TR}}^2(f)},
\end{equation}
which can be rewritten in the same form as the conventional model for wireless communication channels:
\begin{equation}\label{eq:in-out-wireless}
 V_{\textrm{R}}(f) = H(f)\, V_{\text{T}}(f) + W(f).   
\end{equation}
In (\ref{eq:in-out-wireless}), $V_{\textrm{T}}(f)$ is the frequency domain representation of the real pass-band signal, $v(t)$, to be transmitted over the channel. $H(f)$ represents the transfer function of the channel and $W(f)$ is the Fourier transform of the noise $w(t)$ given by:
\begin{subequations}\label{siso-wireless-channel-quantities}
    \begin{align}
    &H(f) = \frac{\beta\,R_{\text{in}}\,Z_{\text{RT}}(f)}{\big(R_{\text{in}}+Z_{\text{R}}(f)\big)\,\big(R+Z_{\text{T}}(f)\big) - Z_{\text{TR}}^2(f)}, \label{eq:siso-channel}\\
    &W(f) = \widetilde{V}_{\text{N,LNA}}(f) + \beta\,R_{\text{in}}\,\frac{Z_{\text{RT}}(f)\, \widetilde{V}_{\textrm{N,T}}(f) + (R+Z_{\text{T}}(f))\,\widetilde{V}_{\textrm{N,R}}(f)}{\big(R_{\text{in}}+Z_{\text{R}}(f)\big)\,\big(R+Z_{\text{T}}(f)\big) - Z_{\text{TR}}^2(f)},\label{eq:siso-noise}
    \end{align}
\end{subequations}
wherein the self and mutual (NF and FF) impedances, i.e., diagonal and off-diagonal entries of $\mathbf{Z}_{\text{SISO}}$, can be derived using the electromagnetic coupling model as will be shown in Section~\ref{sec:NF-FF}. One can also notice that the noise in (\ref{eq:siso-noise}) depends on self/mutual impedances, unlike the conventional assumption of independent additive Gaussian noise. The derived SISO model in (\ref{eq:in-out-wireless}) and (\ref{siso-wireless-channel-quantities}) requires the expressions of the NF and FF mutual impedances, namely, $\mathbf{Z}_{\text{SISO}}^{\textrm{NF}}$ and $\mathbf{Z}_{\text{SISO}}^{\textrm{NF}}$, to fully characterize the SISO communication circuit model given in Fig.~\ref{fig:siso-system-model-nomn}. The analytical expressions of these two impedance matrices will be established in Section~\ref{sec:NF-FF}.
\subsection{Circuit-theoretic models for near- and far-field SISO communications}\label{subsec:NF-siso-model}
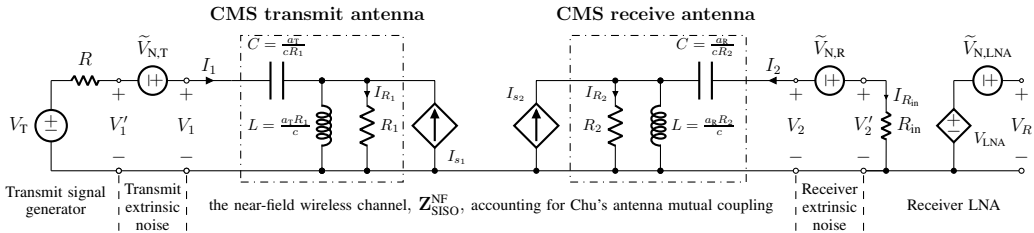
\begin{figure}[h!]
\centering
 \begin{circuitikz}[american voltages, american currents, scale=0.6, every node/.style={transform shape}]
\draw (-4.5,2) to[/tikz/circuitikz/bipoles/length=33pt, V, l_=$V_{\textrm{T}}$] (-4.5,0);
\draw (-4.5,2) to[/tikz/circuitikz/bipoles/length=20pt,R,l=$R$] (-3,2);
\draw (-4.5,0) to[short,-o] (-3,0);
\draw  (-1.5, 2) to[/tikz/circuitikz/bipoles/length=30pt,V, label=\mbox{}, l_=$\widetilde{V}_{\textrm{N,T}}$, o-o] (-3, 2);
\draw (-3,0) to[short, o-o] (-1.5,0);
\draw (-3, 2) to [open,v=$V_1^\prime$] (-3,0);
\draw (-1.5,0) to[short,o-*] (1.5,0);
\draw (1.5,0) to[short,*-*] (2.5,0);
\draw (1.5,2) to[short,*-*] (2.5,2);
\draw (-1.5,2) to [short,i>^=$I_1$, o-] (-0.5,2);
\draw (-0.5,2) to[C, label=\mbox{\small{$C=\frac{a_{\text{T}}}{c R_1}$}}, -*] (1.5,2);
\draw (1.5,2)  to[L, l_=\mbox{\small{$L=\frac{a_{\text{T}} R_1}{c}$}}, *-*] (1.5,0);
\draw (2.5,2) to[/tikz/circuitikz/bipoles/length=30pt,R, i>^=\mbox{\small{$I_{R_1}$}}, l=\mbox{\small{$R_1$}}, *-*] (2.5,0);
\draw (4,0) to[cI, label=\mbox{}, *-] (4,2)
(2.5,2) to [short,-] (4,2)
(-1.5, 2) to [open,v=$V_1$] (-1.5,0)
(2.5,0) to[short, *-*] (6.25,0)
(6.25,0) to[cI, label=\mbox{}] (6.25,2)
(6.25, 2) to[short,-*] (8,2)
(6.25, 0) to[short,-*] (8,0)
(8,2)  to[/tikz/circuitikz/bipoles/length=30pt,R, l_=\mbox{\small{$R_2$}}, i>_=\mbox{\small{$I_{R_2}$}}, *-*] (8,0)
(8, 2) to[short,-*] (9,2)
(8, 0) to[short,-*] (9,0)
(9,0)  to[L, l_=\mbox{\small{$L=\frac{a_{\text{R}} R_2}{c}$}}, *-*] (9,2)
(9,0) to[short,-] (11,0)
(9,2) to[C, label=\mbox{\small{$C=\frac{a_{\text{R}}}{c R_2}$}}, ] (11,2)
(12, 2) to [open,v=$V_2$] (12,0)
(12,2) to [short,i>_=$I_2$] (11,2)
(12,0) -- (11,0)
(13.5, 2) to [open,v=$V_2^\prime$] (13.5,0);
\draw  (13.5,2) to[/tikz/circuitikz/bipoles/length=30pt,V, label=\mbox{}, l_=$\widetilde{V}_{\textrm{N,R}}$, -o] (12,2);
\draw (13.5,2) to[short, o-] (14,2);
\draw (14,2)  to[/tikz/circuitikz/bipoles/length=20pt,R, l^=\mbox{{$R_{{\textrm{in}}}$}}, i>^=\mbox{{$I_{R_{{\textrm{in}}}}$}}, -*] (14,0);
\draw (12,0) to[short,o-*] (14,0);
\draw (13.5,0) to[short,o-*] (15.5,0);
\draw  (15.5, 2) to[/tikz/circuitikz/bipoles/length=30pt,cV, label=\mbox{}] (15.5, 0);
\draw  (17,2) to[/tikz/circuitikz/bipoles/length=30pt,V, label=\mbox{},l_=$\widetilde{V}_{\textrm{N,LNA}}$, o-] (15.5,2);
\draw (15.5,0) to[short,-o] (17,0);
\draw (17,2) to [open,v={{$V_R$} }] (17,0);
\node[] at (5.8,1.7) {\small{$I_{s_2}$}};
\node[] at (4.5,0.3) {\small{$I_{s_1}$}};
\node[] at (16.3,0.7) {\small{$V_{\text{LNA}}$}};
\node[] at (1.4,3.5) {$\mathbf{CMS ~transmit~ antenna}$};
\node[] at (8.9,3.5) {$\mathbf{CMS ~receive~ antenna}$};
\draw[dashdotted] (-0.3,-0.25) rectangle +(3.6,3.25);
\draw[dashdotted] (7,-0.25) rectangle +(3.9,3.25);
\draw [dashed] (-1.5, -0.1) to (-1.5, -1.5);
\draw [dashed] (12, -0.1) to (12, -1.5);
\draw [dashed] (13.5, -0.1) to (13.5, -1.5);
\draw [dashed] (-3, -0.1) to (-3, -1.5);
\node[] at (-4.4,-0.5) {$\small{\textrm{Transmit signal}}$};
\node[] at (-4.4,-0.9) {$\small{\textrm{generator}}$};
\node[] at (5.25,-0.75) {$\small{\text{the near-field wireless channel},\,\mathbf{Z}_\text{SISO}^{\textrm{NF}}, \,\textrm{accounting for Chu's antenna mutual coupling}}$};
\node[] at (12.75,-0.4) {$\small{\textrm{Receiver}}$};
\node[] at (12.75,-0.8) {$\small{\textrm{extrinsic}}$};
\node[] at (12.75,-1.2) {$\small{\textrm{noise}}$};
\node[] at (15.5,-0.75) {$\small{\textrm{Receiver LNA}}$};
\node[] at (-2.25,-0.4) {$\small{\textrm{Transmit}}$};
\node[] at (-2.25,-0.8) {$\small{\textrm{extrinsic}}$};
\node[] at (-2.25,-1.2) {$\small{\textrm{noise}}$};
\end{circuitikz}
\caption{Near-field SISO communication model of Fig.~\ref{fig:siso-system-model-nomn} wherein $\mathbf{Z}_{\text{SISO}}$ was replaced by the NF channel $\mathbf{Z}_{\text{SISO}}^{\textrm{NF}}$ composed of two electrical Chu antennas from Fig.~\ref{fig:tm1}. The transmit\big/receive antennas involve controlled current sources $I_{s_1}$\big/$I_{s_2}$ by $I_{R_2}$\big/$I_{R_1}$ which model their mutual electromagnetic influence in the NF region.}
\label{fig:siso-near-field-system-model-nomn}
\end{figure}

The SISO channel model established in (\ref{eq:in-out-wireless}) and (\ref{siso-wireless-channel-quantities}) is not tailored to Chu's CMS antennas. Its equivalent circuit in Fig.~\ref{fig:siso-system-model-nomn} must include the circuit of Chu's CMS antennas from Fig.~\ref{fig:tm1}. By doing so, this will fully characterize its NF and FF electromagnetic radiation properties when the mutual impedances of Chu's CMS antennas are specified. Using Chu's CMS antenna modelled in Fig.~\ref{fig:tm1} at both the transmitter and the receiver, the SISO model in Fig.~\ref{fig:siso-system-model-nomn} gives a near-field circuit-theoretic model depicted in Fig.~\ref{fig:siso-near-field-system-model-nomn}. There, the two current-dependent sources, $I_{s_1}$ and $I_{s_2}$, account for the transmitter-receiver and receiver-transmitter reaction currents when the transmitter and the receiver are in close proximity (i.e., near-field).

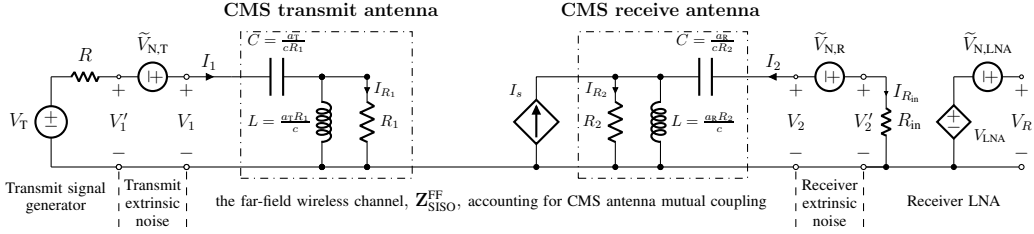
\begin{figure}[h!]
\centering
 \begin{circuitikz}[american voltages, american currents, scale=0.6, every node/.style={transform shape}]
\draw (-4.5,2) to[/tikz/circuitikz/bipoles/length=33pt, V, l_=$V_{\textrm{T}}$] (-4.5,0);
\draw (-4.5,2) to[/tikz/circuitikz/bipoles/length=20pt,R,l=$R$] (-3,2);
\draw (-4.5,0) to[short,-o] (-3,0);
\draw  (-1.5, 2) to[/tikz/circuitikz/bipoles/length=30pt,V, label=\mbox{}, l_=$\widetilde{V}_{\textrm{N,T}}$, o-o] (-3, 2);
\draw (-3,0) to[short, o-o] (-1.5,0);
\draw (-3, 2) to [open,v=$V_1^\prime$] (-3,0);
\draw (-1.5,0) to[short,o-*] (1.5,0);
\draw (1.5,0) to[short,*-*] (2.5,0);
\draw (1.5,2) to[short,*-*] (2.5,2);
\draw (-1.5,2) to [short,i>^=$I_1$, o-] (-0.5,2);
\draw (-0.5,2) to[C, label=\mbox{\small{$C=\frac{a_{\text{T}}}{c R_1}$}}, -*] (1.5,2);
\draw (1.5,2)  to[L, l_=\mbox{\small{$L=\frac{a_{\text{T}} R_1}{c}$}}, *-*] (1.5,0);
\draw (2.5,2) to[/tikz/circuitikz/bipoles/length=30pt,R, i>^=\mbox{\small{$I_{R_1}$}}, l=\mbox{\small{$R_1$}}, *-*] (2.5,0);
\draw (-1.5, 2) to [open,v=$V_1$] (-1.5,0)
(2.5,0) to[short, *-*] (6.25,0)
(6.25,0) to[cI, label=\mbox{}] (6.25,2)
(6.25, 2) to[short,-*] (8,2)
(6.25, 0) to[short,-*] (8,0)
(8,2)  to[/tikz/circuitikz/bipoles/length=30pt,R, l_=\mbox{\small{$R_2$}}, i>_=\mbox{\small{$I_{R_2}$}}, *-*] (8,0)
(8, 2) to[short,-*] (9,2)
(8, 0) to[short,-*] (9,0)
(9,0)  to[L, l_=\mbox{\small{$L=\frac{a_{\text{R}} R_2}{c}$}}, *-*] (9,2)
(9,0) to[short,-] (11,0)
(9,2) to[C, label=\mbox{\small{$C=\frac{a_{\text{R}}}{c R_2}$}}, ] (11,2)
(12, 2) to [open,v=$V_2$] (12,0)
(12,2) to [short,i>_=$I_2$] (11,2)
(12,0) -- (11,0)
(13.5, 2) to [open,v=$V_2^\prime$] (13.5,0);
\draw  (13.5,2) to[/tikz/circuitikz/bipoles/length=30pt,V, label=\mbox{}, l_=$\widetilde{V}_{\textrm{N,R}}$, -o] (12,2);
\draw (13.5,2) to[short, o-] (14,2);
\draw (14,2)  to[/tikz/circuitikz/bipoles/length=20pt,R, l^=\mbox{{$R_{{\textrm{in}}}$}}, i>^=\mbox{{$I_{R_{{\textrm{in}}}}$}}, -*] (14,0);
\draw (12,0) to[short,o-*] (14,0);
\draw (13.5,0) to[short,o-*] (15.5,0);
\draw  (15.5, 2) to[/tikz/circuitikz/bipoles/length=30pt,cV, label=\mbox{}] (15.5, 0);
\draw  (17,2) to[/tikz/circuitikz/bipoles/length=30pt,V, label=\mbox{},l_=$\widetilde{V}_{\textrm{N,LNA}}$, o-] (15.5,2);
\draw (15.5,0) to[short,-o] (17,0);
\draw (17,2) to [open,v={{$V_R$} }] (17,0);
\node[] at (5.8,1.7) {\small{$I_{s}$}};
\node[] at (16.3,0.7) {\small{$V_{\text{LNA}}$}};
\node[] at (1.7,3.5) {$\mathbf{CMS ~transmit~ antenna}$};
\node[] at (9.,3.5) {$\mathbf{CMS ~receive~ antenna}$};
\draw[dashdotted] (-0.3,-0.25) rectangle +(3.8,3.25);
\draw[dashdotted] (7.18,-0.25) rectangle +(3.75,3.25);
\draw [dashed] (-1.5, -0.1) to (-1.5, -1.5);
\draw [dashed] (12, -0.1) to (12, -1.5);
\draw [dashed] (13.5, -0.1) to (13.5, -1.5);
\draw [dashed] (-3, -0.1) to (-3, -1.5);
\node[] at (-4.4,-0.5) {$\small{\textrm{Transmit signal}}$};
\node[] at (-4.4,-0.9) {$\small{\textrm{generator}}$};
\node[] at (5.25,-0.75) {$\small{\text{the far-field wireless channel},\,\mathbf{Z}_\text{SISO}^{\textrm{FF}}, \,\textrm{accounting for CMS antenna mutual coupling}}$};
\node[] at (12.75,-0.4) {$\small{\textrm{Receiver}}$};
\node[] at (12.75,-0.8) {$\small{\textrm{extrinsic}}$};
\node[] at (12.75,-1.2) {$\small{\textrm{noise}}$};
\node[] at (15.5,-0.75) {$\small{\textrm{Receiver LNA}}$};
\node[] at (-2.25,-0.4) {$\small{\textrm{Transmit}}$};
\node[] at (-2.25,-0.8) {$\small{\textrm{extrinsic}}$};
\node[] at (-2.25,-1.2) {$\small{\textrm{noise}}$};
\end{circuitikz}
\caption{Far-field SISO communication model of Fig.~\ref{fig:siso-system-model-nomn} wherein $\mathbf{Z}_{\text{SISO}}$ was replaced by the FF channel $\mathbf{Z}_{\text{SISO}}^{\textrm{FF}}$ composed of two electrical Chu antennas from Fig.~\ref{fig:tm1}. Only the receive antenna involves a controlled current source $I_{s}$ by $I_{R_1}$ to model the electromagnetic influence of the transmit antenna on the receive antenna in the FF region.}
\label{fig:siso-far-field-system-model-nomn}
\end{figure}

\noindent Unlike the near-field circuit model in Fig.~\ref{fig:siso-near-field-system-model-nomn}, the far-field model illustrated in Fig. \ref{fig:siso-far-field-system-model-nomn} does not involve the receiver-transmitter reaction current $I_{s_1}$ since the electrical properties of the receiver do not influence those of the transmitter, i.e., $Z_{\text{TR}} = 0$.

\noindent As will be shown in Section \ref{sec:NF-FF}, this difference affects the computation of the mutual impedances of Chu's CMS antennas only. The self-impedances, $Z_{\text{T}}$ and $Z_{\text{R}}$, remain unchanged since they do not depend on the separating distance between the transmitter and the receiver. Furthermore, in accordance with Figs.~\ref{fig:siso-near-field-system-model-nomn} and \ref{fig:siso-far-field-system-model-nomn}, both $Z_{\text{T}}$ and $Z_{\text{R}}$ are obtained from (\ref{eq:Z1-chu}) as the self-impedances of the transmit and receive Chu's antennas with sizes $a_{\text{T}}$ and $a_{\text{R}}$ and resistances $R_1$ and $R_2$, respectively.

\section{Computation of the near- and far-field impedance matrices}\label{sec:NF-FF}
In this section, we derive the mutual impedances involved in the NF and FF SISO channel impedance matrices, $\mathbf{Z}_{\text{SISO}}^{\textrm{NF}}$ and $\mathbf{Z}_{\text{SISO}}^{\textrm{FF}}$. In particular, the NF mutual impedances are derived using the induced EMF method owing to the Chu-Hertz equivalence established in Appendix~\ref{appendix:self-mutual-impedance-Chu}. The FF mutual impedance, however, is obtained based on the Friis' equation.

\subsection{Computation of the near-field mutual impedances}
\subsubsection{Induced EMF analysis and the Chu-Hertz equivalence}
The induced EMF method consists in studying the electromotive forces that are induced on the antenna structure under a known current distribution. Given an incident field, $E_{\text {in}}$, impinging on a receive antenna of aperture $\mathcal{A}$, the induced open-circuit voltage, $V_{\mathrm{oc}}$, at its terminals is given by \cite{jordan1968electromagnetic}
\begin{equation}\label{eq:induced-EMF}
    V_{\mathrm{oc}}=-\frac{1}{I_0} \int_{\mathcal{A}} E_{\text {in}}(\ell) \,I(\ell) \,\text{d}\ell,
\end{equation}
where $I_0$ is the input port current when the antenna is transmitting. The explicit knowledge of the current distribution $I(\ell)$ in (\ref{eq:induced-EMF}) renders the induced EMF method incompatible with Chu's CMS antennas whose radiated EM fields (described in Appendix~\ref{appendix:Chu-radiated-EM-fields}) are known outside their encompassing spheres only. To overcome this limitation, we resort to the equivalence theorem and establish a relationship between the mutual impedances of two Hertz dipoles, $\left(Z^{\text{Hertz}}_{\text{RT}},Z^{\text{Hertz}}_{\text{TR}}\right)$, and the mutual impedances of two Chu's CMS antennas, $\left(Z^{\text{Chu}}_{\text{RT}},Z^{\text{Chu}}_{\text{TR}}\right)$.
To that end, we show in Appendix~\ref{appendix:Chu-Hertz-equivalence} (cf. Result~\ref{result:result1} below) that the EM fields and the radiation resistance of a Chu antenna \big($\mathbf{E}_{\textrm{Chu}}$, $\mathbf{H}_{\textrm{Chu}}$\big) are equivalent to \big($\mathbf{E}_{\textrm{Hertz}}$, $\mathbf{H}_{\textrm{Hertz}}$\big) of a Hertz dipole when the complex coefficient $A_1$ of the $\text{TM}_1$ mode of radiation in (\ref{appendix-eq:V-I-TM1}) is appropriately chosen.
\begin{result}\label{result:result1}
  The radiation of a Hertz dipole $\Re\big[Z^{\textrm{Hertz}}_{\text{SISO}}\big]$ having a uniform current distribution, $I$, and a Chu's antenna $\Re\big[Z^{\textrm{Chu}}_{\text{SISO}}\big]$ can be made equal by adequately choosing the mode coefficient, $A_1$, of the $\textrm{TM}_1$ mode pertaining to the Chu's electric antenna is
\begin{equation}\label{eq:A1-equivalence-condition-Chu-Hertz-main-text}
    A_1 = j\,\frac{I k^2 c}{4\pi f} \sqrt{\frac{3\, \Re\big[Z^{\textrm{Hertz}}_{\text{SISO}}\big]}{2\pi\eta}} ~\iff~ \Re\big[Z^{\textrm{Chu}}_\text{SISO}\big] = \Re\big[Z^{\textrm{Hertz}}_\text{SISO}\big].
\end{equation}
\end{result}
\begin{proof}
See Appendix~\ref{appendix:Chu-Hertz-equivalence}.
\end{proof}
\noindent Using the equivalent radiated power, which is a direct consequence of the radiation resistance equivalence in (\ref{eq:A1-equivalence-condition-Chu-Hertz-main-text}), we derive the following relationship between the mutual impedances of two Chu’s antennas and two Hertz dipoles:\vspace{-0.2cm}
\begin{result}\label{result:equivalence-chu-hertz}
The mutual impedance equivalence between two Hertz dipoles and two Chu's antennas is given by:
\end{result}
\vspace{-1.3cm}
\begin{subequations}\label{eq:I-chu-I-hertz-final-main-text}
\begin{align}
    \frac{Z^{\textrm{Hertz}}_{\text{RT}}}{\sqrt{\Re\big[Z^{\textrm{Hertz}}_{\text{T}}\big] \, \Re\big[Z^{\textrm{Hertz}}_{\text{R}}\big]}} &=\frac{Z^{\textrm{Chu}}_{\text{RT}}}{\sqrt{\Re\big[Z^{\textrm{Chu}}_{\text{T}}\big] \, \Re\big[Z^{\textrm{Chu}}_{\text{R}}\big]}},\label{eq:I-chu-I-hertz-final-main-text-a}\\
    \frac{Z^{\textrm{Hertz}}_{\text{TR}}}{\sqrt{\Re\big[Z^{\textrm{Hertz}}_{\text{T}}\big] \, \Re\big[Z^{\textrm{Hertz}}_{\text{R}}\big]}} &=\frac{Z^{\textrm{Chu}}_{\text{TR}}}{\sqrt{\Re\big[Z^{\textrm{Chu}}_{\text{T}}\big] \, \Re\big[Z^{\textrm{Chu}}_{\text{R}}\big]}}.\label{eq:I-chu-I-hertz-final-main-text-b}
\end{align}
\end{subequations}

\begin{proof}
See Appendix~\ref{appendix:self-mutual-impedance-Chu}.
\end{proof}
\noindent It is therefore enough to apply the induced EMF method on two Hertz dipoles and then deduce the mutual impedance between two Chu's antennas using Result~\ref{result:equivalence-chu-hertz}.

\subsubsection{Near-field mutual impedance calculation}
\begin{figure}[h!]
    \centering
    \includegraphics[scale=0.9]{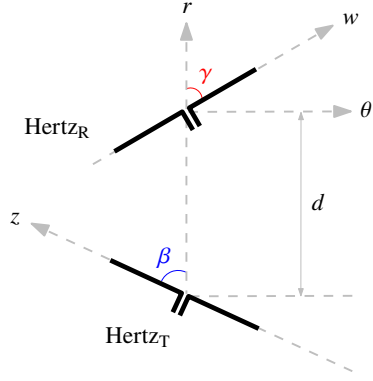}
    \caption{Transmit and receive Hertz dipoles $\text{Hertz}_\text{T}$\big/$\text{Hertz}_\text{R}$ of length $\text{d}\ell_\text{T}$\big/$\text{d}\ell_\text{R}$ in the same plane, arbitrarily oriented in free space, and separated (in close proximity) by a distance $d$ [m].}
    \label{fig:two-hertz-antennas}
    \vspace{-0.5cm}
\end{figure}

We consider a pair of transmit and receive Hertz dipoles in the same plane. The dipoles are separated by a distance $d$, aligned with respect to (w.r.t.) their axes $w$ and $z$, respectively, and arbitrarily rotated with angles $\beta$ and $\gamma$ w.r.t. their connecting axis $r$ as depicted in Fig.~\ref{fig:two-hertz-antennas}.

The mutual impedances between the two Hertz dipoles, $\text{Hertz}_\text{T}$ and $\text{Hertz}_\text{R}$, are given by (see Appendix~\ref{appendix:self-mutual-impedance-SISO-Hertz}):
\begin{equation}\label{eq:impedance-Z-21-final-hertz-main-text}
    \begin{aligned}[b]
        Z^{\text{Hertz}}_{\text{RT}} = Z^{\text{Hertz}}_{\text{TR}} &=-\frac{3\,k_0^2\,c^2}{4\pi^2f^2}\,\sqrt{\Re\big[Z^{\textrm{Hertz}}_{\text{T}}\big]\,\Re\big[Z^{\textrm{Hertz}}_\text{R}\big]} \Bigg[\frac{1}{2}\,\sin(\beta)\,\sin(\gamma)\bigg(\frac{1}{j\,k_0\,d} + \frac{1}{(j\,k_0\,d)^2} \\ &\hspace{1cm}+\frac{1}{(j\,k_0\,d)^3}\bigg)
        + \cos(\gamma)\,\cos(\beta)\bigg(\frac{1}{(j\,k_0\,d)^2} + \frac{1}{(j\,k_0\,d)^3}\bigg)\Bigg]\,e^{-j\,k_0\,d}.
    \end{aligned}
\end{equation}
The fact that the mutual impedances are equal is a consequence of  the Lorentz reciprocity theorem. Indeed, the latter states that the reaction of the transmit antenna's electromagnetic fields to the current in the receive antenna is equal to the reaction of the receive antenna's electromagnetic fields to the current in the transmit antenna. Moreover, when the two antennas are of the same type/size, the self-impedances must be equal. Therefore, the near-field communication channel is represented by a 2$\times$2 symmetric impedance matrix $\mathbf{Z}_{\text{SISO}}^{\textrm{NF}}$ whose diagonal entries are equal only when $a_{\textrm{T}} = a_{\textrm{R}}$.

\noindent Using (\ref{eq:impedance-Z-21-final-hertz-main-text}) and the Chu-Hertz mutual impedance equivalence established in Result~\ref{result:equivalence-chu-hertz}, the mutual impedances between two Chu antennas are obtained as:
\begin{equation}\label{eq:impedance-Z-21-final-chu-main-text}
    \begin{aligned}[b]
        &Z^{\text{Chu}}_{\text{RT}} = Z^{\text{Chu}}_{\text{TR}} =-\frac{3\,k_0^2\,c^2}{4\pi^2f^2}\,\sqrt{\Re\big[Z^{\textrm{Chu}}_{\text{T}}\big]\,\Re\big[Z^{\textrm{Chu}}_\text{R}\big]} \Bigg[\frac{1}{2}\,\sin(\beta)\,\sin(\gamma)\bigg(\frac{1}{j\,k_0\,d} + \frac{1}{(j\,k_0\,d)^2} +\frac{1}{(j\,k_0\,d)^3}\bigg)\\
        &\hspace{3.5cm}\times\cos(\gamma)\,\cos(\beta)\bigg(\frac{1}{(j\,k_0\,d)^2} + \frac{1}{(j\,k_0\,d)^3}\bigg)\Bigg]\,e^{-j\,k_0\,d}.
    \end{aligned}
\end{equation}
\noindent Now that we obtained the NF mutual impedances, we turn our focus in the next subsection to the FF case where we derive the FF mutual impedance based on the Friis' transmission equation.

\subsection{Computation of the far-field mutual impedance}
Unlike the computation of the NF mutual impedances where a minimum of EM theory was invoked to model the MC effect, the FF channel model in (\ref{fig:siso-far-field-system-model-nomn}) is governed by the Friis' equation which relates the squared current magnitudes of the transmitter and the receiver as follows:
\begin{equation}\label{eq:friss-current-magnitude}
|I_{s}(f)|^2 = 4\,|I_{R_1}(f)|^2\left(\frac{c}{4{\pi}fd}\right)^2G_\textrm{T}\,G_\text{\textrm{R}}\,\frac{R_1}{R_2}\,\, [\textrm{A}^2].
\end{equation}
In (\ref{eq:friss-current-magnitude}), $d$ is the distance between the transmit and receive antennas which have gains $G_{\textrm{T}}$ and $G_{\textrm{R}}$, respectively. For transmit/receive Chu's CMS antenna, we have that $G_{\textrm{T}}$ and $G_{\textrm{R}}$ are equal to $3/2$ in the equatorial plane \cite[chapter 6]{harrington1961pp}. Using basic circuit analysis and (\ref{fig:siso-far-field-system-model-nomn}), the expression of the FF mutual impedance, $Z_{\mathrm{RT}}^{\textrm{Chu}}(f)$, is expressed as (see Appendix~\ref{appendix:FF-mutual-impedance}):
\begin{equation}\label{eq:FF-mutual-impedance}
    Z_{\mathrm{RT}}^{\textrm{Chu}}(f) = \left( \frac{j\,2\pi\,f\,a_{\textrm{T}}}{c + j\,2\pi\,f\,a_{\textrm{T}}} \right)\frac{c}{2{\pi}fd}\,\sqrt{\frac{G_\text{T}\,G_\text{R}\,R_1}{R_2}}\,\frac{j\,2\pi\,f\,R_2}{j\,2\pi\,f + \frac{c}{a_{\textrm{R}}}}\,[\Omega].
\end{equation}

\noindent The other mutual impedance $Z_{\mathrm{TR}}^{\textrm{Chu}}(f)$ represents the proportionality coefficient between the transmit voltage, $V_1(f)$, and the receive current, $I_2(f)$. Since antennas are reciprocal, we have $Z_{\textrm{RT}}(f) = Z_{\textrm{TR}}(f)$. Moreover, the signal attenuation in the far-field region between the transmitter and the receiver is very large, i.e.:
\begin{equation}
    \big|Z_{\textrm{TR}}^{\textrm{Chu}}(f)\big| ~=~ \big|Z_{\textrm{RT}}^{\textrm{Chu}}(f)\big| ~\ll~ \big|Z_{\textrm{T}}^{\textrm{Chu}}(f)\big|.
\end{equation}
This justifies the so-called ``unilateral approximation'' \cite{ivrlavc2010toward} which stipulates that in the far-field region one has $\big| Z_{\textrm{TR}}^{\textrm{Chu}}(f) \big| \approx 0$. As a consequence, the FF impedance matrix $\mathbf{Z}_{\text{SISO}}^{\textrm{FF}}$ is neither symmetric nor diagonal since only the transmitter is driving the electrical properties of the receiver. Now that the input-output relationship in (\ref{siso-wireless-channel-quantities}) is fully characterized for both NF and FF SISO communications, we are ready to find the achievable rate under transmit and receive antenna size constraints.

\section{Achievable rate optimization methodology}\label{sec:achiveable-rate-optimization}

The maximum achievable rate is the largest mutual information between the input and the output of the communication channel. By inspecting Figs.~\ref{fig:siso-near-field-system-model-nomn} and \ref{fig:siso-far-field-system-model-nomn}, it is seen that the transmit voltage waveform $ v_{\textrm{T}} (t) $ is under full control of the system designer. Unlike traditional purely statistical channel models (e.g. AWGN channels), however, it is now possible to also optimize the NF and FF impedance matrices, $\mathbf{Z}_\text{SISO}^{\textrm{NF}}$ and $\mathbf{Z}_\text{SISO}^{\textrm{FF}}$, to improve the achievable rate. The maximum achievable rate is thus given by:
\begin{subequations}\label{eq:capacity-and-lowerbound}
\begin{align}
C &= \max_{\mathbf{Z}_{\text{SISO}},\,\mathbb{P}_{v_\text{T}}} I(v_{\text{T}}(t);v_{\text{R}}(t))\label{eq:mutual-info-definition}\\ 
&=\max_{\mathbf{Z}_{\text{SISO}},\,\mathbb{P}_{v_\text{T}}}\int_{0}^{\infty}{\log_2\left(1+\frac{P_{\rm t}(f)\,|H(f)|^2}{N(f)}\right)\textrm{d}f}\\
&\approx \max_{\mathbb{P}_{v_\text{T}}}\int_{0}^{\infty}{\log_2\left(1+\frac{P_{\rm t}(f)\,|H(f)|^2}{N(f)}\right)\bigg|_{\mathbf{Z}_{\text{SISO}}=\mathbf{Z}^{\text{Chu}}_{\text{SISO}}}\,\,\textrm{d}f}\,\,[{\textrm{bits/s}}],\label{eq:capacity}
\end{align}
\end{subequations}
where the last approximation follows from the fact that we consider the ``optimal'' (i.e. broadest) transmit/receive antennas for a fixed size\footnote{This is why we only use the terminology ``maximum achievable rate'' instead of ``capacity'' throughout the paper.}. In (\ref{eq:mutual-info-definition}), $I(v_{\text{T}}(t);v_{\text{R}}(t))$ is the mutual information between the two random processes representing the input and output voltages/signals of the communication system~\cite{gel1957computation,gallager1968information}. Moreover, $\mathbb{P}_{v_\text{T}}$ is the probability measure on the space of input/generator voltages, $v_\text{T}(t)$, which for any finite set of time instants $\{t_1,t_2,\ldots,t_n\}$, specifies the following joint cumulative distribution function:
\begin{equation}
\mathbb{P}_{v_\text{T}}[v_\text{T}(t_1)\leq v_1,v_\text{T}(t_2)\leq v_2,\ldots v_\text{T}(t_n)\leq v_n],~~ \forall(v_1,v_2,\ldots,v_n)\in \mathbb{R}^n.
\end{equation}
 In designing the probability law of the generator, we suppose that the expected per-frequency power, $P_{\rm t}(f)$, satisfies:
 \begin{equation}\label{transmit_PSD_constraint}
     P_{\rm t}(f) \leq P_{\textrm{max}},~~\forall\,f,
 \end{equation}
 where $P_{\textrm{max}}$ is the maximum power that the generator can supply due to regulatory restrictions or hardware constraints. In the sequel, we study the impact of the physical size constraints at both the transmitter and the receiver by restricting the volumes encompassing their antennas to be of finite radii, namely $a_{\text{T}}$ and $a_{\text{R}}$, respectively.
\subsection{Maximum achievable rate of near-field SISO wireless channels}
The maximum achievable rate given in (\ref{eq:AWGN_Capacity}) can now be evaluated using the channel response $H(f)$ established in (\ref{eq:siso-channel}) and the PSD derived for the noise $W(f)$ obtained in (\ref{eq:siso-noise}). To this end, we introduce the following two quantities:
\begin{subequations}\label{eq:variables-X-Y}
    \begin{align}
    X(f) &= \big|Z_{\text{RT}}^{\text{Chu}}(f)\big|^2\,\Re{\big[Z_{\text{T}}^{\text{Chu}}(f)\big]} + \left(R+\xbar{Z^{\text{Chu}}_{\text{T}}(f)}\right)\,Z_{\text{TR}}^{\text{Chu}}(f)\,\Re{\big[Z^{\text{Chu}}_{\text{RT}}(f)\big]} \nonumber \\
    & \hspace{1cm}+ \left(R+Z_{\text{T}}^{\text{Chu}}(f)\right)\,\xbar{Z^{\text{Chu}}_{\text{TR}}(f)}\,\,\Re{\big[Z^{\text{Chu}}_{\text{TR}}(f)\big]} + \big|R + Z^{\text{Chu}}_{\text{T}}(f)\big|^2\,\Re{\big[Z^{\text{Chu}}_{\text{R}}(f)\big]},\\
    Y(f) & =  \Big|\left(R_{\textrm{in}}+Z^{\text{Chu}}_{\text{R}}(f)\right)\left(R+Z^{\text{Chu}}_{\text{T}}(f)\right)-\left(Z^{\text{Chu}}_{\text{TR}}(f)\right)^2\Big|^2.
    \end{align}
\end{subequations}

\noindent By employing (\ref{eq:variables-X-Y}) and the noise auto/cross-correlation properties in (\ref{eq:cross-correlation-zero})--(\ref{eq:cross-correlation-NF}), we show that:\\
$i)$ The square magnitude of the channel response $H(f)$ is expressed as: 
\begin{equation}
    |H(f)|^2 = \frac{\beta^2\,R_{\text{in}}^2\,\big|Z^{\text{Chu}}_{\text{RT}}(f)\big|^2}{Y(f)}\label{eq:channel-power}.
\end{equation}
$ii)$ The PSD of the noise is given by:
\begin{equation}\label{eq:noise-power}
    N(f) = k_\text{b}\,T\,\frac{R_{\textrm{in}}}{R}\left[(N_{\textrm{f}}-1) +  \beta^2\,R_{\textrm{in}} \,\frac{X(f)}{Y(f)}\right]. 
\end{equation}

\noindent With (\ref{eq:channel-power}) and (\ref{eq:noise-power}), the achievable rate in (\ref{eq:AWGN_Capacity}) is now fully characterized as function of the self/mutual impedances of Chu's CMS antennas. The band-limited transmit PSD $P_t(f)$ is the only remaining variable subject to optimization, which will be the subject of the next subsection.
 \subsection{Achievable rate with optimal transmit power allocation}
 Under a total power budget, $P_{\textrm{max}}$, the maximum achievable rate in (\ref{eq:AWGN_Capacity}) can be optimized w.r.t. the transmit PSD, $P_{\rm t}(f)$, thereby leading to the following power allocation (PA) problem:
 \begin{equation}\label{eq:power-allocation-optimization}
 \begin{aligned}
     C_{\textrm{PA}} &= \max_{P_{\rm t}(f)}\int_{0}^{\infty}{\log_2\left(1+\frac{P_{\rm t}(f)\,|H(f)|^2}{N(f)}\right)\textrm{d}f}\,\,[{\textrm{bits/s}}],\\
     &\textrm{subject to} \quad \int_{f_c-\frac{W}{2}}^{f_c+\frac{W}{2}}{P_{\rm t}(f)\,\textrm{d}f} \leq P_{\textrm{max}},
\end{aligned}
 \end{equation}
 
\noindent where $W$ is the absolute bandwidth and $f_c$ is the center frequency of $W$. The solution of (\ref{eq:power-allocation-optimization}) can be found iteratively by the well-know frequency-domain water filling procedure \cite{gallager1968information}:
  \begin{equation}\label{eq:power-allocation-optimization-solution}
 \begin{aligned}
    P_{\textrm{t}}^{\tiny{\starletfill}}(f) = 
    \left\{
        \begin{array}{ll}
            P_{\textrm{max}}\,\max\left(0,\frac{1}{\gamma_0} - \frac{1}{\gamma(f)}\right) &,\, \gamma(f) \geq \gamma_0 \\
            0 & ,\, \gamma(f) < \gamma_0.
        \end{array}
    \right.
\end{aligned}
 \end{equation}
 Here, $\gamma(f) = |H(f)|^2/N(f)$ and $\gamma_0 = \mu\,P_{\textrm{max}}\,\log(2)$ with $\mu$ being the Lagrange multiplier of the Lagrangian function associated to (\ref{eq:power-allocation-optimization}). By letting $P_{\textrm{t}}(f) = P_{\textrm{t}}^{\tiny{\starletfill}}(f)$, the maximum achievable rate under optimal power allocation, $C_{\textrm{PA}}$, becomes:
 \begin{equation}\label{capacity-PA}
     C_{\textrm{PA}} = \int_{0}^{\infty}{\max\left(0,\,\log_2\left(\frac{\gamma(f)}{\gamma_0}\right)\right)\textrm{d}f}\,\,[{\textrm{bits/s}}].
 \end{equation}
 The threshold value $\gamma_0$ can be obtained by numerically integrating the following power constraint:
 \begin{equation}
     \int_{0}^{\infty}{
     \frac{P^{\tiny{\starletfill}}_{\rm t}(f)}{P_{\textrm{max}}}}\,\textrm{d}f = \int_{0}^{\infty}{
     \left(\frac{1}{\gamma_0} - \frac{1}{\gamma(f)}\right)}\,\textrm{d}f = 1.
 \end{equation}
Using the established expression of the achievable rate under antenna size constraints for both NF and FF scenarios, we now examine the effect of the transmit and receive antenna sizes on the overall performance of SISO communications.
 
 \section{Numerical results and discussions}\label{sec:results}
 In this section, we present numerical results for the maximum achievable rate of SISO communications under antenna size constraints at both the transmitter and the receiver by inspecting the behavior of the following metrics:
 \begin{itemize}[leftmargin=*]
     \item the SNR as a function of the frequency with $\text{SNR}(f)=P_{\rm t}(f)\,|H(f)|^2 / N(f)$,
     \item the maximum achievable rate in (\ref{eq:AWGN_Capacity}) under transmit and receive antenna size constraints with uniform power allocation,
     \item the maximum achievable rate in (\ref{capacity-PA}) under optimal power allocation.
 \end{itemize}

\begin{figure}[h!]
    \begin{subfigure}{0.45\textwidth}
        \centering
        \vspace{1cm}
        \includegraphics[scale=0.9]{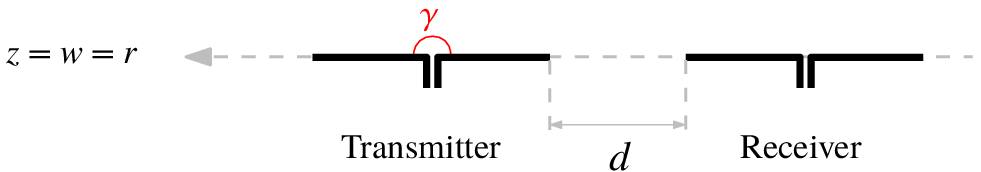}
        \vspace{0.05cm}
        \caption{colinear configuration}
        \label{fig:siso-colinear-config}
    \end{subfigure}
\hfill
    \begin{subfigure}{.45\textwidth}
        \centering
        \includegraphics[scale=0.9]{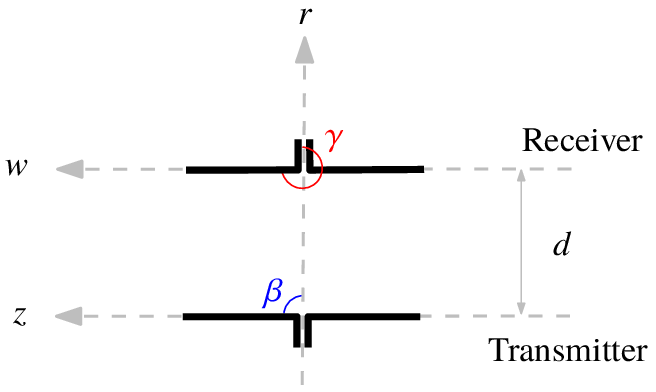}
        \caption{parallel configuration}
        \label{fig:siso-parallel-config}
    \end{subfigure}
\caption{Two special relative orientation between the transmit and receive antennas of the general case shown in Fig.~\ref{fig:two-hertz-antennas}: (a) colinear configuration with $\gamma=\pi$ and $\beta =0$, (b) parallel configuration with $\gamma=\frac{3\pi}{2}$ and $\beta=\frac{\pi}{2}$.}
    \label{fig:siso-geometry-configurations}
\end{figure}

\noindent We also examine the performance under the following two configurations:
\begin{enumerate}
   \item[$a)$] colinear configuration as depicted in Fig.~\protect\ref{fig:siso-colinear-config} where the axes of transmit and receive antennas are colinear,
   \item[$b)$] parallel configuration as depicted in Fig.~ \ref{fig:siso-parallel-config} where the axes of transmit and receive antennas are parallel.
\end{enumerate}

\noindent In both scenarios, the Chu's sphere encompassing the transmit (resp. receive) antenna has a radius $a_\text{T}$ (resp. $a_\text{R}$) with $a_\text{T} + a_\text{R} \leq d$. The latter condition ensures that the transmit and receive spheres do not overlap in the near-field region so that the equivalence theorem can still be used to find the corresponding mutual impedances.
\subsection{SNR under uniform power control}
\begin{figure}[h!]
    \centering
    \includegraphics[scale=0.5]{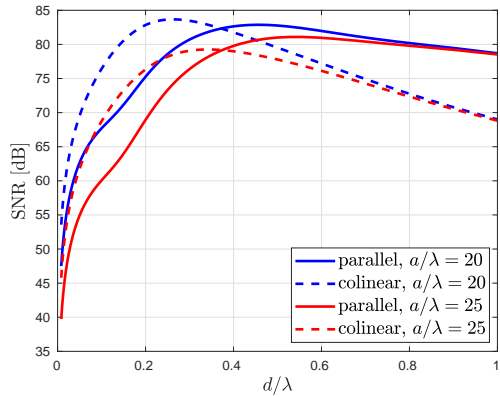}
    \caption{$\textrm{SNR}$ as a function of the electrical distance/separation, $d/\lambda$, for both the parallel and colinear configurations and two different antenna sizes, $a$, with $P_{\textrm{max}} = 10\, [\text{mW}]$ and $d = 4\,[\text{mm}]$.} 
    \label{fig:SNR}
\end{figure}

We first plot in Fig.~\ref{fig:SNR} the SNR as a function of the electrical distance/separation, $d/\lambda$, for two antenna sizes, $a/\lambda_{p} \in \{20, 25\}$ with $(a_{\textrm{T}}, a_{\textrm{R}}) = (a,a)$, for both the colinear and parallel configurations. Here, $\lambda_{p}$ is the peak wavelength, i.e., the wavelength calculated at the highest \textrm{SNR}. 
In (\ref{eq:noise-power}), we set the noise factor to $N_\text{f} = 5\,\textrm{dB}$ and the noise temperature to $T = 300\,[\textrm{K}]$.
Fig.~\ref{fig:SNR} reveals that the colinear antenna configuration yields higher \textrm{SNR} than the parallel configuration as long as $d/\lambda \lessapprox 0.38$. Thereafter, the SNR decreases quickly and becomes much smaller than that of the parallel configuration. The threshold distance $d\approx 0.38\,\lambda$ can be interpreted as the limit between the NF and FF regions. In fact, when antennas are closely spaced, most of the communication happens through the radial/NF component of the electric field $E_r$ which is aligned along the colinear axis and vanishes in the FF region. It should be noticed that when $d\approx\lambda$ the degradation of the $\textrm{SNR}$ due to finite antenna sizes almost vanishes. This can be attributed to the fact that the antenna size relative to the wavelength increases thereby making the antennas not electrically small anymore, i.e., they do not store a lot of near-field reactive energy.

\begin{figure}[h!]
    \centering
    \includegraphics[scale=0.5]{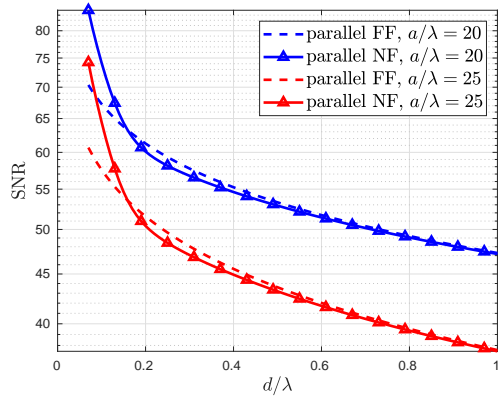}
    \caption{$\textrm{SNR}$ as a function of the electrical distance/separation, $d/\lambda$, for the parallel configuration and two different antenna sizes, $a$, with $f_c = 1\,[\text{GHz}]$ and $P_{\textrm{max}} = 10\, [\text{mW}]$.} 
    \label{fig:SNR-NF-vs-FF}
\end{figure}

\noindent We also compare in Fig.~\ref{fig:SNR-NF-vs-FF} the SNR using the exact mutual impedance in (\ref{eq:impedance-Z-21-final-chu-main-text}) against the SNR based on the FF mutual impedance in (\ref{eq:FF-mutual-impedance}). We observe that the FF approximation becomes reasonably accurate starting from $d\approx0.5\,\lambda$. This confirms the widely adopted half-wavelength spacing choice in 4G/5G antenna design for a fixed antenna size.

\subsection{Maximum achievable rate under uniform power control}

\begin{figure}[h!]
    \centering
    \subcaptionbox{Reactive near-field region at $d=0.15\,\lambda$.\label{fig:achievable-rate-vs-size-d-0.3}}{{
    \includegraphics[scale=0.33]{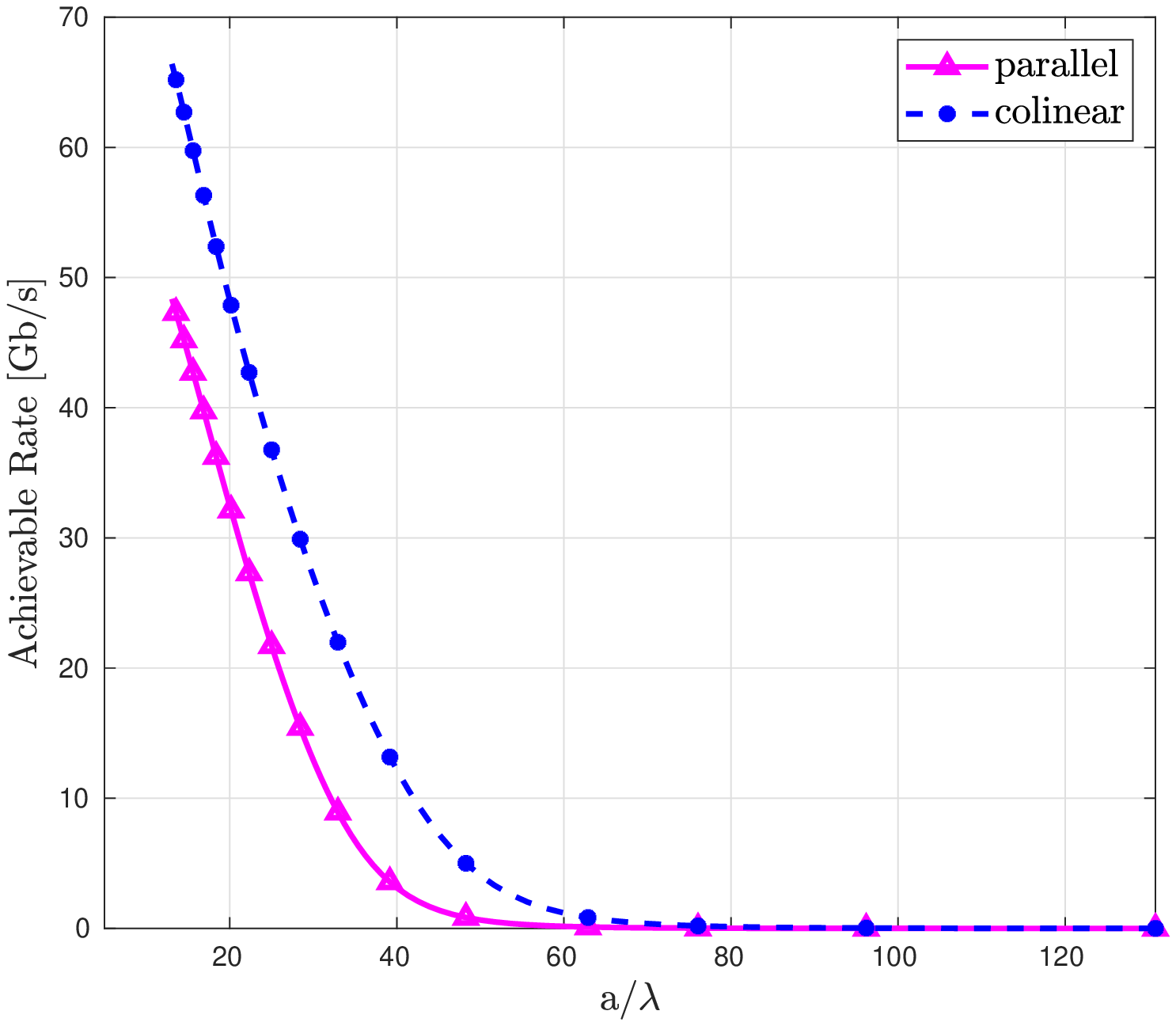} 
    }}%
    \subcaptionbox{Radiative near-field region at $d=0.45\,\lambda$.\label{fig:achievable-rate-vs-size-d-0.4}}{{
    \includegraphics[scale=0.33]{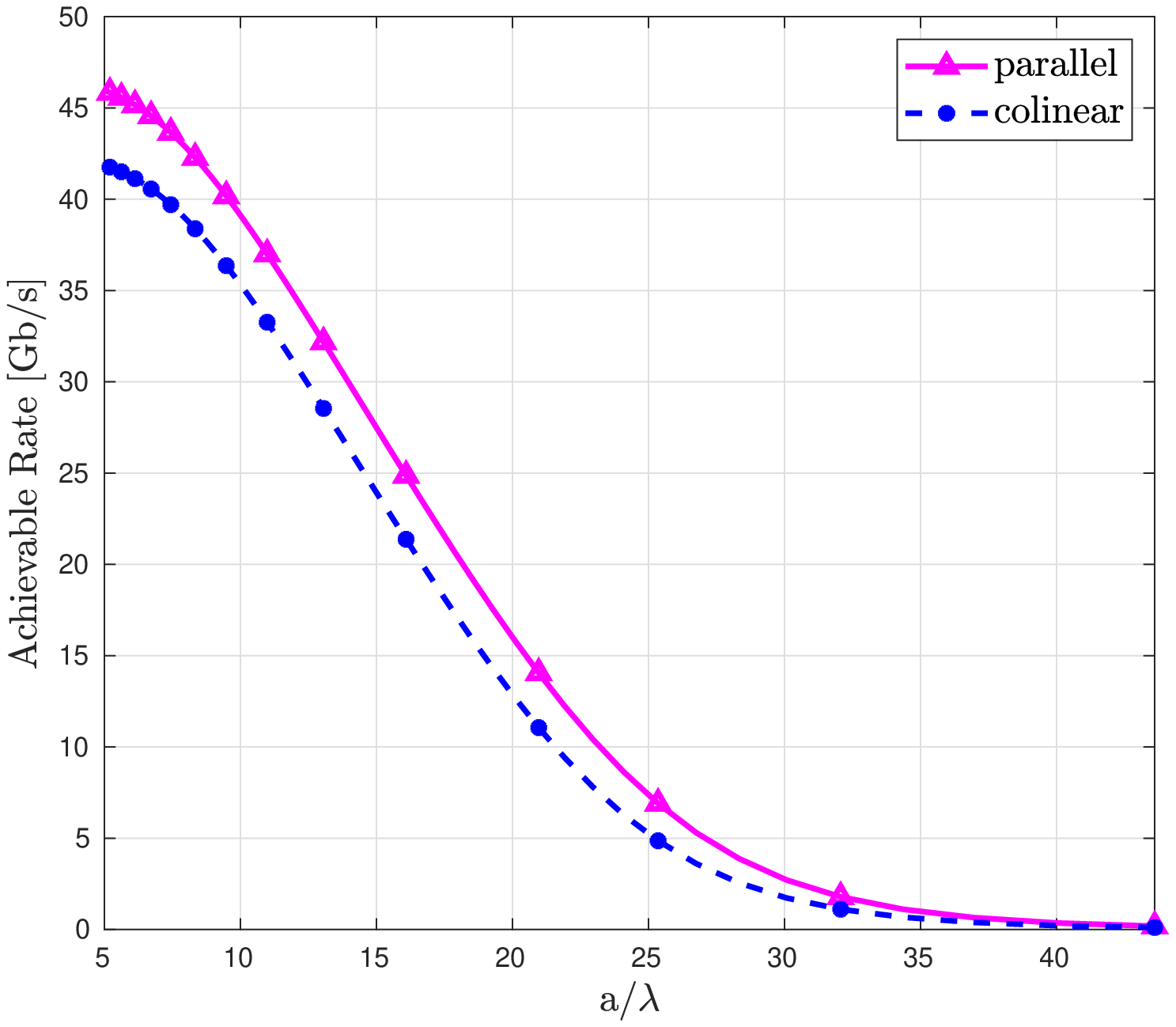} 
    }}
    \hspace{1cm}
    \subcaptionbox{Far-field region at $d=2\,\lambda$.\label{fig:achievable-rate-vs-size-FF}}{{
    \includegraphics[scale=0.33]{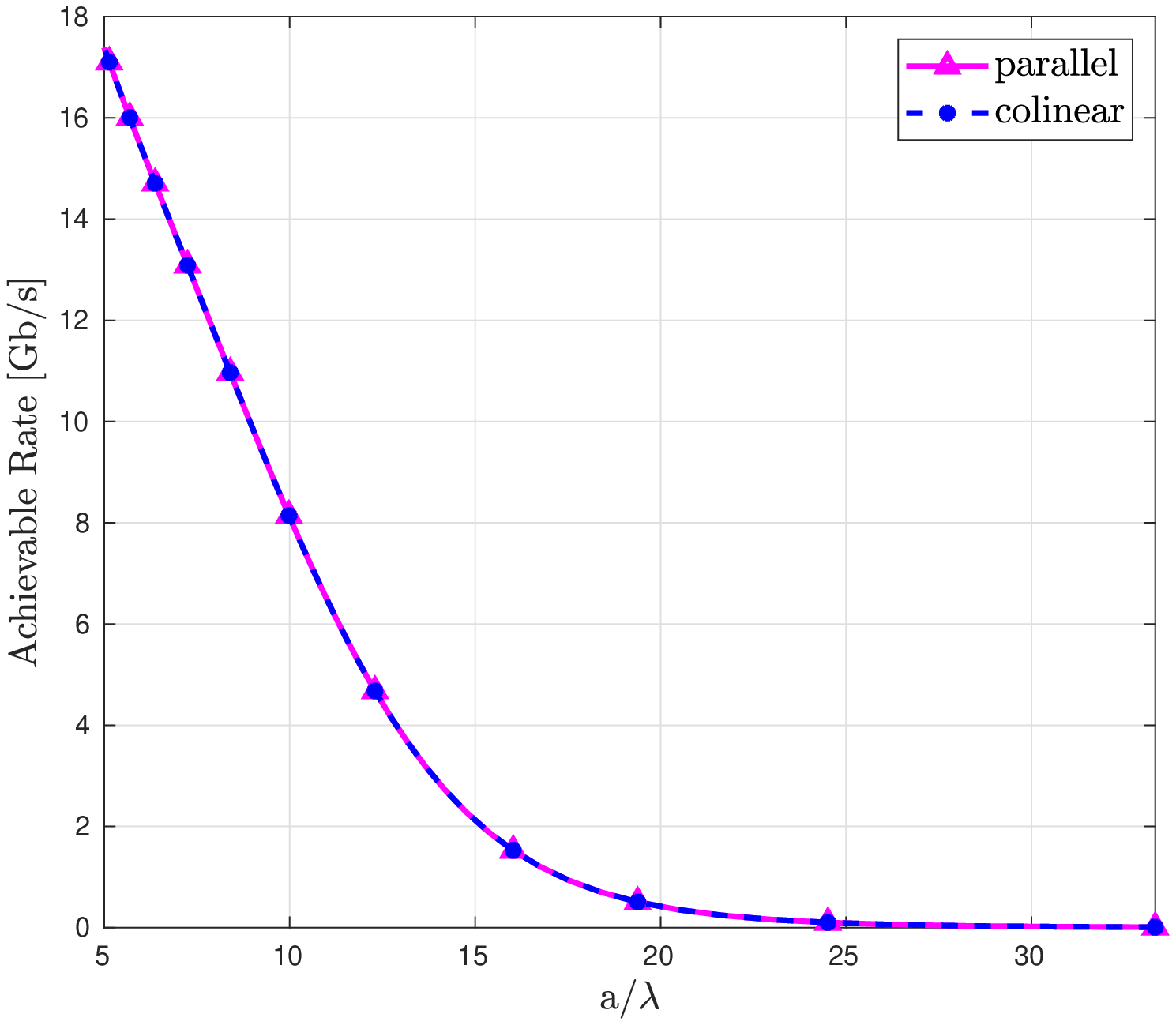}
    }}%
    \caption{Plots of the achievable rate as a function of the the electrical distance/separation, $a/\lambda$, for the parallel and colinear configurations with $f_c = 25\, [\text{GHz}]$, $W = 0.2\,f_c$, $a=\lambda/20$, and $P_{\textrm{max}} = 10\, [\text{mW}]$ in (a) the reactive near-field region, (b) the radiative near-field region, and (c) the far-field region.}
    \label{fig:achievable-rate-vs-size} 
\end{figure}

In Fig.~\ref{fig:achievable-rate-vs-size}, we vary the electrical size $\lambda/a$ and plot the maximum achievable rate in (\ref{eq:AWGN_Capacity}) 
with the channel response in (\ref{eq:channel-power}) and the noise PSD in (\ref{eq:noise-power}).
The carrier frequency is set to $f_c = 25 \,[\textrm{GHz}]$ and the bandwidth is taken to be $W = 0.2\,f_c$. In Fig.~\ref{fig:achievable-rate-vs-size}, we distinguish three different regimes depending on the antenna separation. These are the reactive NF region depicted in Fig.~\protect\ref{fig:achievable-rate-vs-size-d-0.3}, the radiative NF region (a.k.a. Fresnel region) depicted in Fig.~\protect\ref{fig:achievable-rate-vs-size-d-0.4}, and the FF region depicted in Fig.~\protect\ref{fig:achievable-rate-vs-size-FF}.
As seen there, the colinear antenna configuration yields a higher achievable rate in the reactive NF region due to the strong radial component of the electric field $E_r$. In contrast, in the radiative NF region the parallel configuration yields a slightly higher $\textrm{SNR}$ than the colinear configuration. In the FF region, both configurations exhibit the same achievable rate since the contribution of the field $E_r$ is negligible. It is also seen that the three regimes admit different achievable rate behaviors, which could not be simply attributed to the difference in the $\textrm{SNR}$ values.

\begin{figure}[h!]
\begin{subfigure}{.5\textwidth}
  \centering
  \includegraphics[width=.8\linewidth]{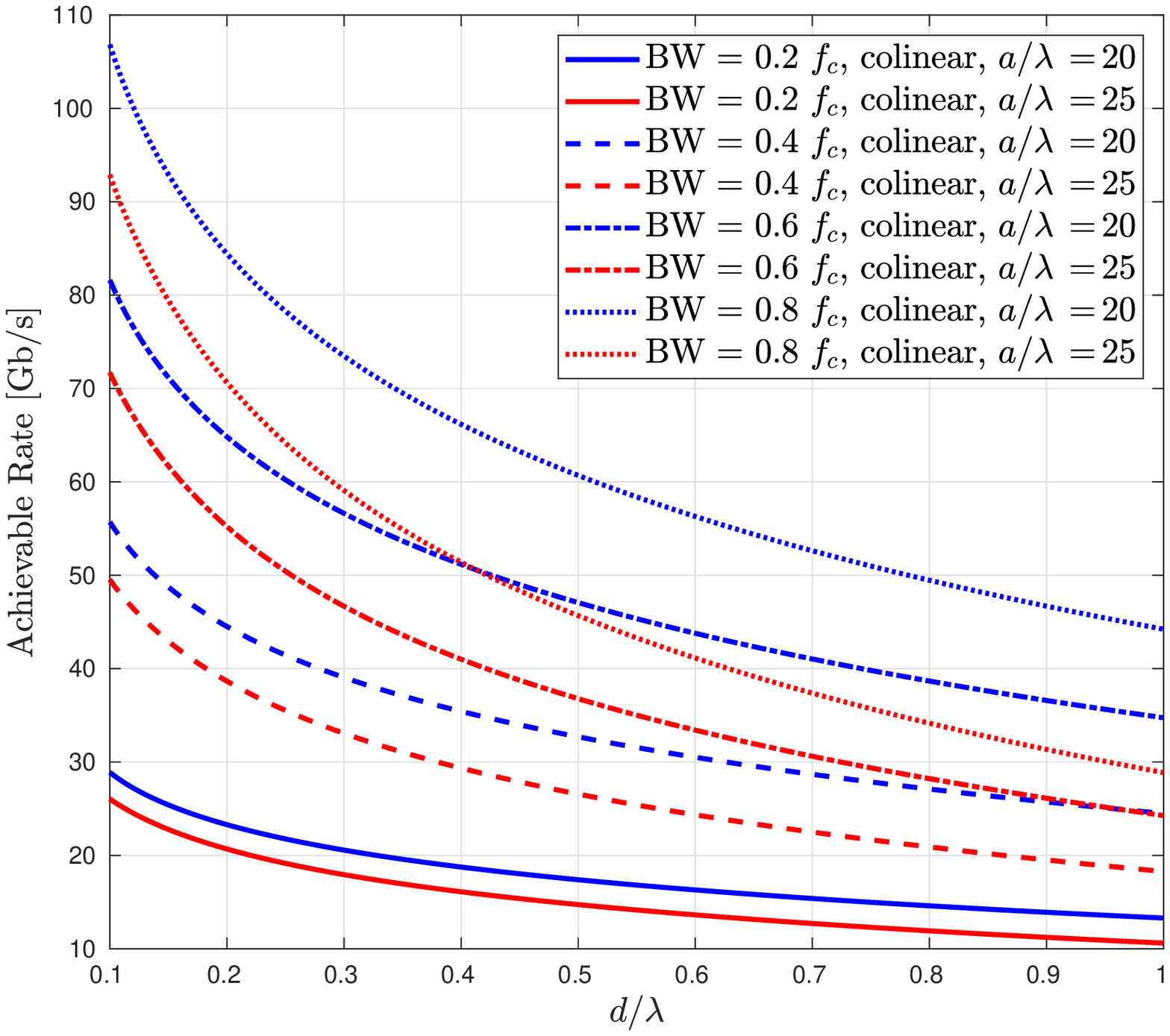}
  \vspace{-0.2cm}
  \caption{Near-field: colinear configuration}
  \label{fig:fractionBW-NF-colinear}
\end{subfigure}%
\begin{subfigure}{.5\textwidth}
  \centering
  \includegraphics[width=.8\linewidth]{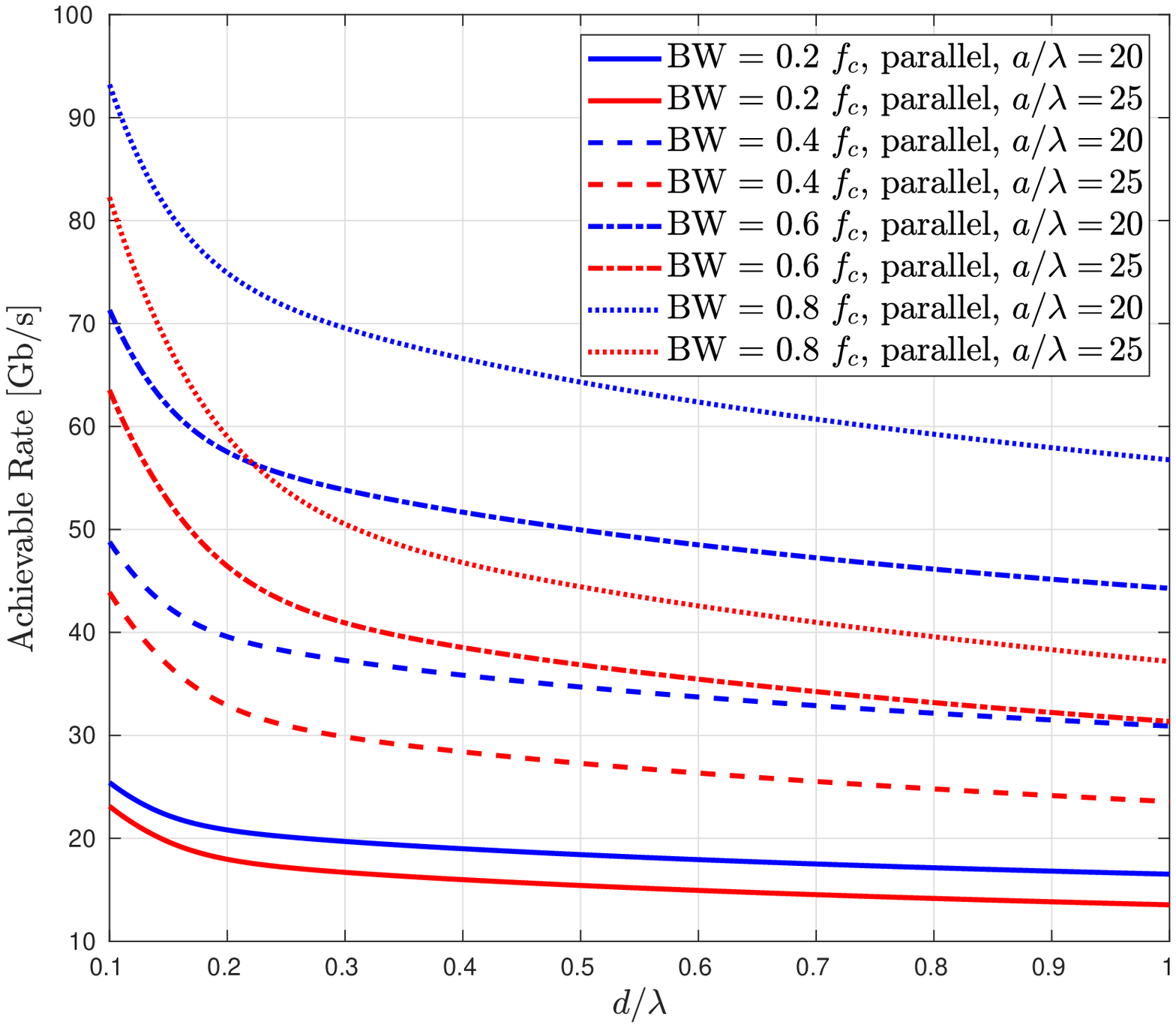}
  \vspace{-0.2cm}
  \caption{Near-field: parallel configuration}
  \label{fig:fractionBW-NF-parallel}
\end{subfigure}
\fboxsep=-\fboxrule\rule{0pt}{3cm}\hspace{4cm}
\begin{subfigure}{.5\textwidth}
  \centering
  \includegraphics[width=.8\linewidth]{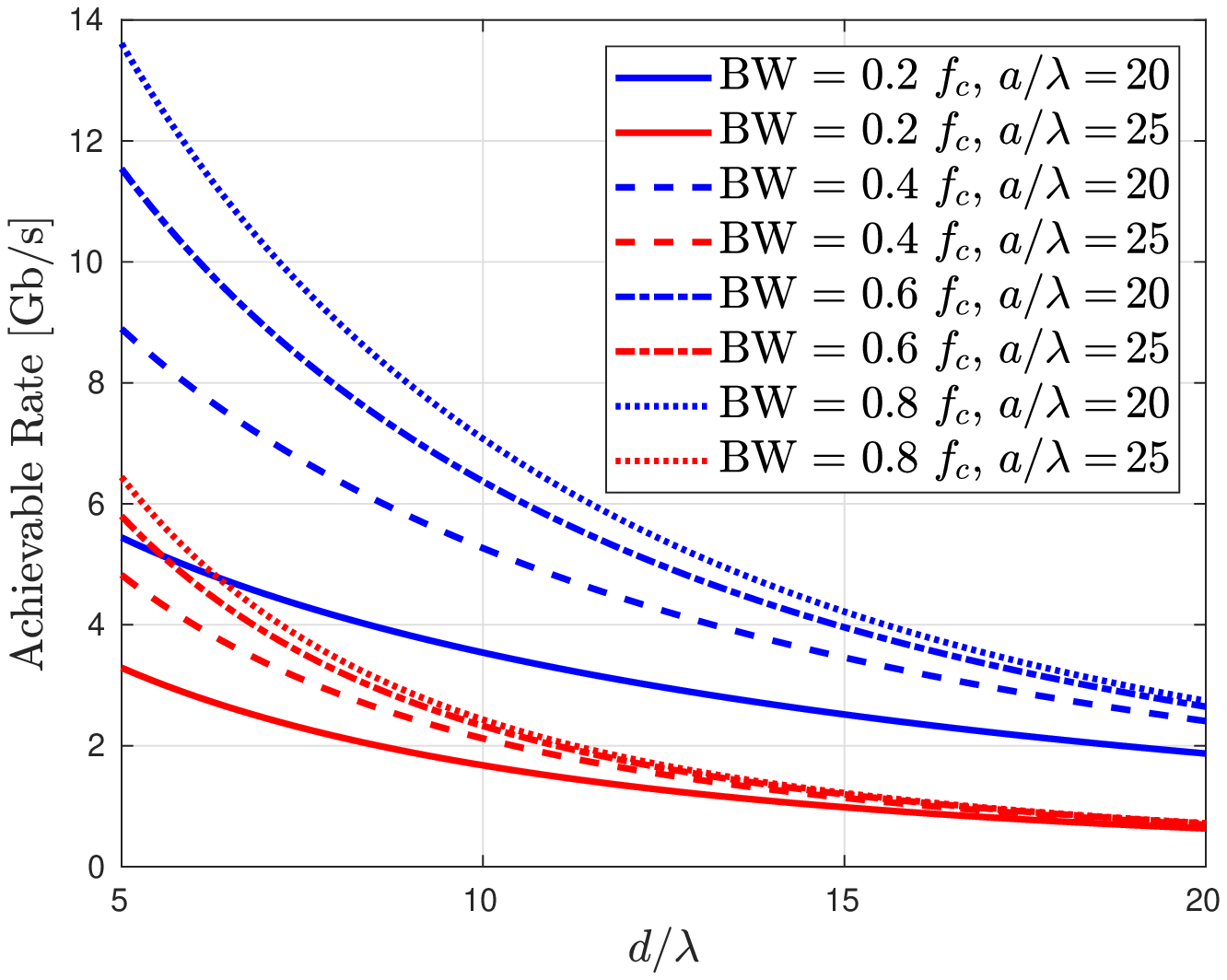}
  \vspace{-0.2cm}
  \caption{Far-field}
  \label{fig:fraction-BW-FF}
\end{subfigure}
\caption{Plots of the achievable rate as a function of $d/\lambda$ at  $f_c = 25\, [\text{GHz}]$ for the near-field region in (a) and (b), and the far-field region in (c) using both colinear and parallel configurations with two antenna sizes and $P_{\textrm{max}}=0.1\,[\textrm{mW}]$. The far-field plot in (c) corresponds to both parallel and colinear configurations.}
\label{fig:fractionBW}
\vspace{-0.3cm}
\end{figure}

\noindent Fig.~\ref{fig:fractionBW} plots the achievable rate as a function of $d/\lambda$ for the same antenna sizes already used in Fig.~\ref{fig:SNR} while varying the bandwidth as a fraction of the carrier frequency $f_c$. There, it is observed that the increase in bandwidth, as well as the antenna size, have a significant effect on the achievable rate in both the parallel and colinear configurations. As already observed in Fig.~\ref{fig:achievable-rate-vs-size}, when the distance is larger than approximately $0.38\,\lambda$, the parallel configuration also yields a higher achievable rate than the colinear configuration and vice versa when $d\lessapprox 0.38\,\lambda$. The FF plot in Fig. \ref{fig:fraction-BW-FF}, for both the parallel and colinear configurations, looks different due to the considered low-SNR regime as opposed to the high-SNR regime in the NF regions shown in Figs. \ref{fig:fractionBW-NF-colinear} and \ref{fig:fractionBW-NF-parallel}. 

\begin{figure}[h!]
    \centering
    \subfloat[Reactive near-field at $d=0.1\,\lambda$]{{
    \includegraphics[scale=0.4]{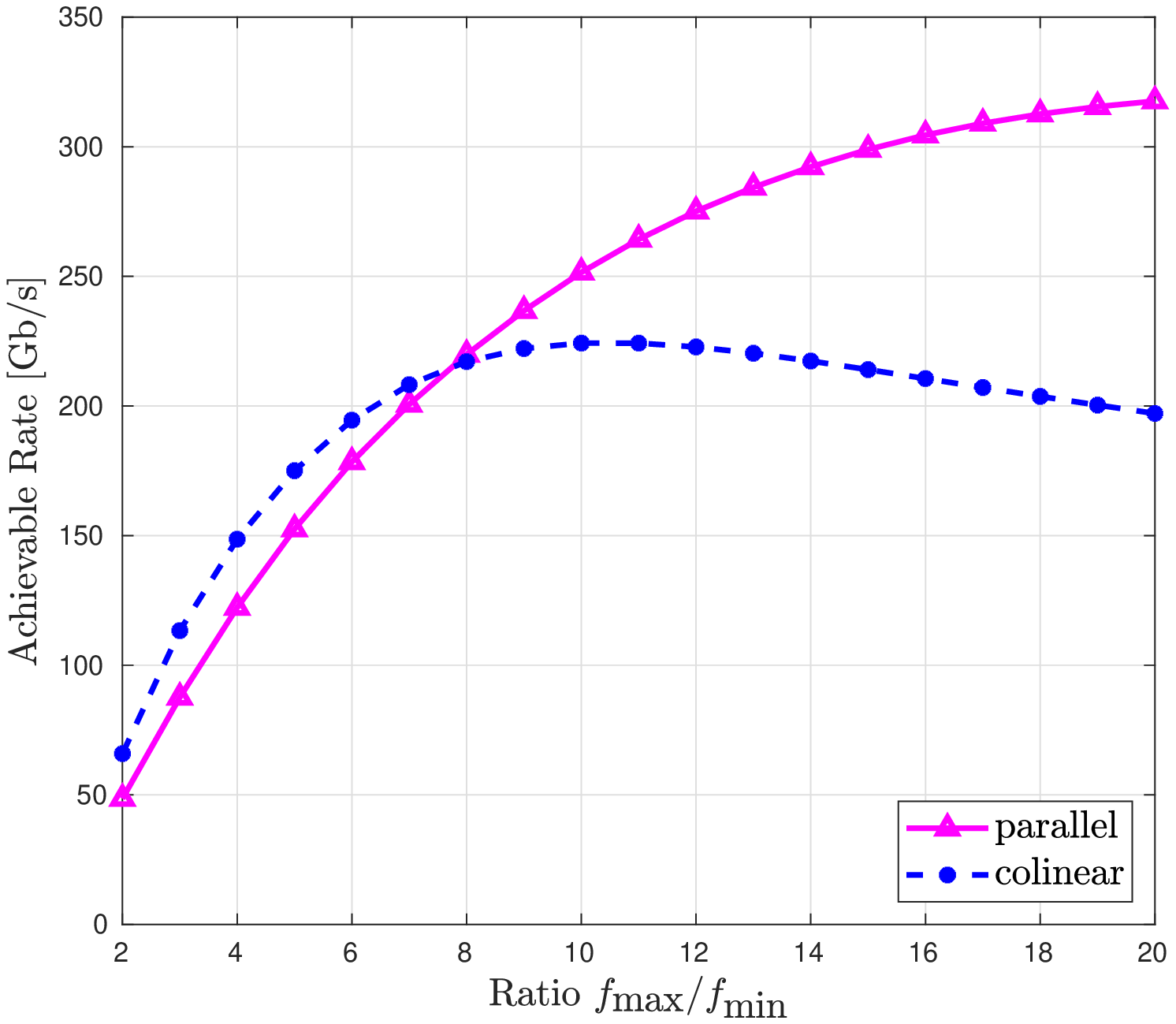} 
    \label{fig:data-rate-NF-0.1}}}
    \vspace{0.001cm}
    \subfloat[Radiative near-field at $d=0.15\,\lambda$]{{
    \includegraphics[scale=0.4]{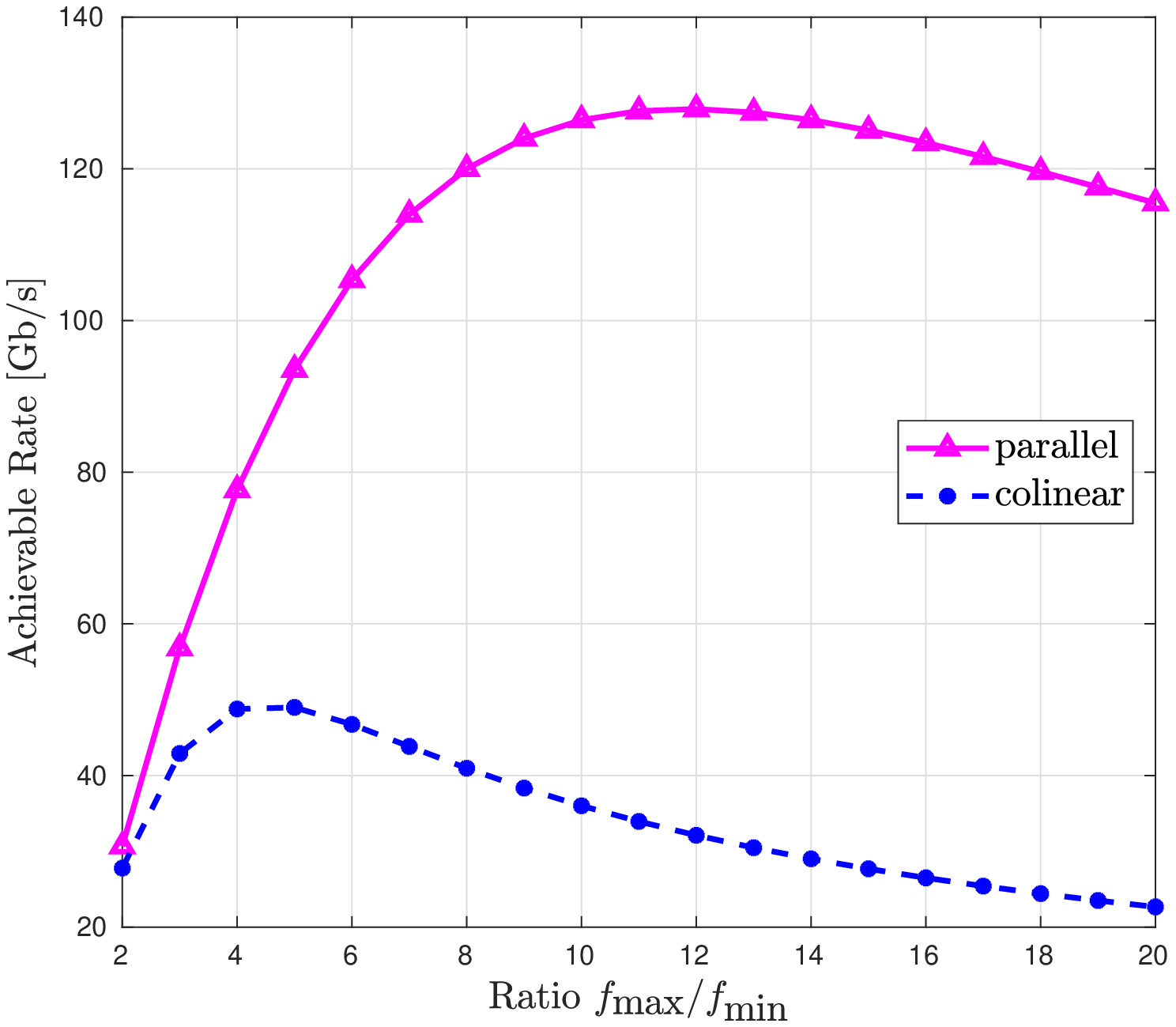} 
    \label{fig:data-rate-NF}}}%
    \subfloat[Far-field at $d=0.5\,\lambda$]{{
    \includegraphics[scale=0.4]{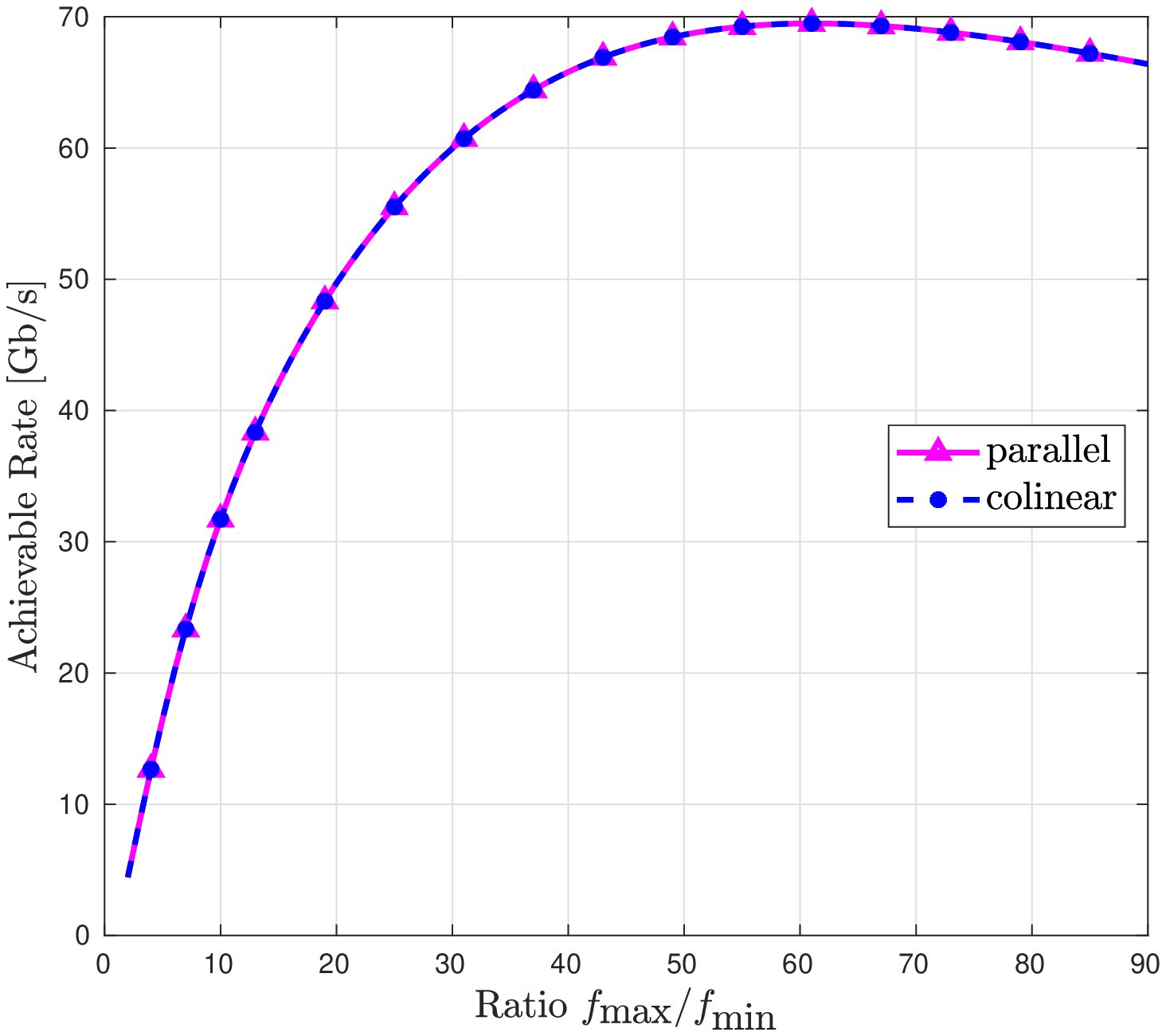} 
    \label{fig:data-rate-FF}}}
    \caption{Plots of the achievable rate as a function of the bandwidth ratio $f_{\text{max}}/f_{\text{min}}$ for the parallel and colinear configurations at $f_\textrm{min} = 5\, [\text{GHz}]$ with $P_{\textrm{max}} = 0.1\, [\text{mW}]$ in the near-field and far-field regions.}
    \label{fig:data-rate}
    \vspace{-0.4cm}
\end{figure}

\noindent In addition, Figs. \ref{fig:fractionBW-NF-colinear} and \ref{fig:fractionBW-NF-parallel} demonstrate that the achievable rate of the colinear configuration (resp., parallel) is higher than that of the parallel (resp., colinear) configuration in the near-field (resp. far-field) region independently of the fraction of the bandwidth, thereby validating the results obtained previously. Motivated by this observation, we inspect more closely the effect of signaling bandwidth by plotting in Fig.~\ref{fig:data-rate} the achievable rate as a function of the so-called bandwidth ratio. The latter is measured as the ratio of the largest frequency, $f_\textrm{max}$, to the smallest frequency, $f_\textrm{min}$, in the frequency band. The results are shown for the reactive NF, radiative NF, and FF regions in Figs.~\ref{fig:data-rate}a, \ref{fig:data-rate}b, and \ref{fig:data-rate}c, respectively.

\noindent In line with the previous numerical results, in the reactive near-field region, the achievable rate is higher for the colinear configuration as long as $f_\textrm{max}/f_\textrm{min} \leq 8$. In the radiative near-field region, however, the parallel configuration provides a higher achievable rate. In the far-field region, the two configurations yield the same performance since the unilateral approximation becomes asymptotically exact as the separation distance goes to infinity. We observe that the optimal signalling bandwidth decreases from the reactive to the radiative near-field region and increases largely in the far-field region. The optimal bandwidth ratio is around $5$ (i.e., $W=15\,[\textrm{GHz}]$) for the colinear configuration in the radiative near-field region all the way up to 60 (i.e., $W=300\,[\textrm{GHz}]$) in the far-field region. It should be noted that the aforementioned values are highly dependent on the transmit power as well as the smallest frequency in the band, and thus must be optimized depending on the situation.
\subsection{Maximum achievable rate under optimal power allocation}

\begin{figure}[h!]
    \centering
    \subfloat[Near-field]{{
    \includegraphics[scale=0.4]{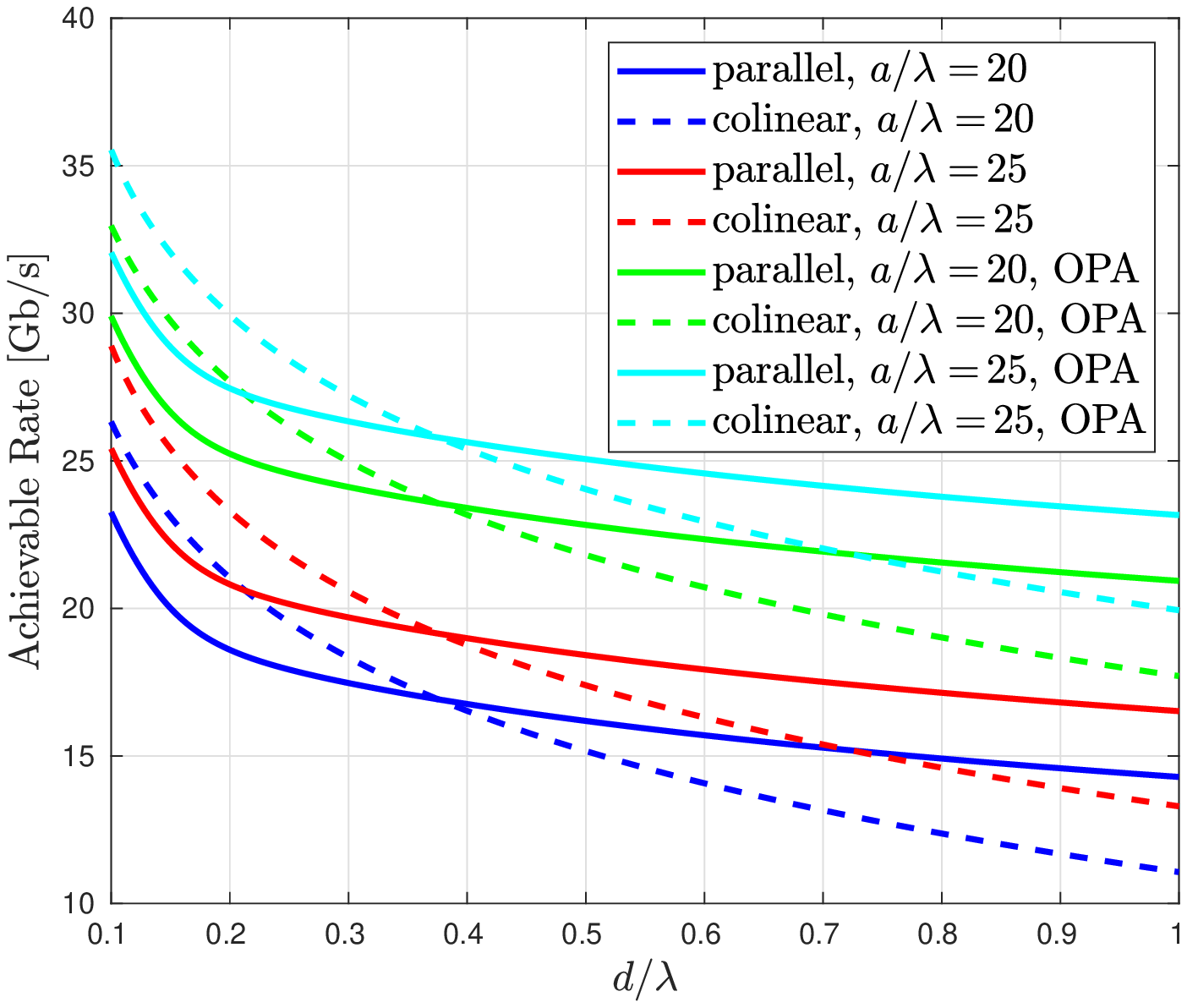} 
    \label{fig:achievable-rate-NF-PA}}}%
    \subfloat[Far-field]{{
    \includegraphics[scale=0.4]{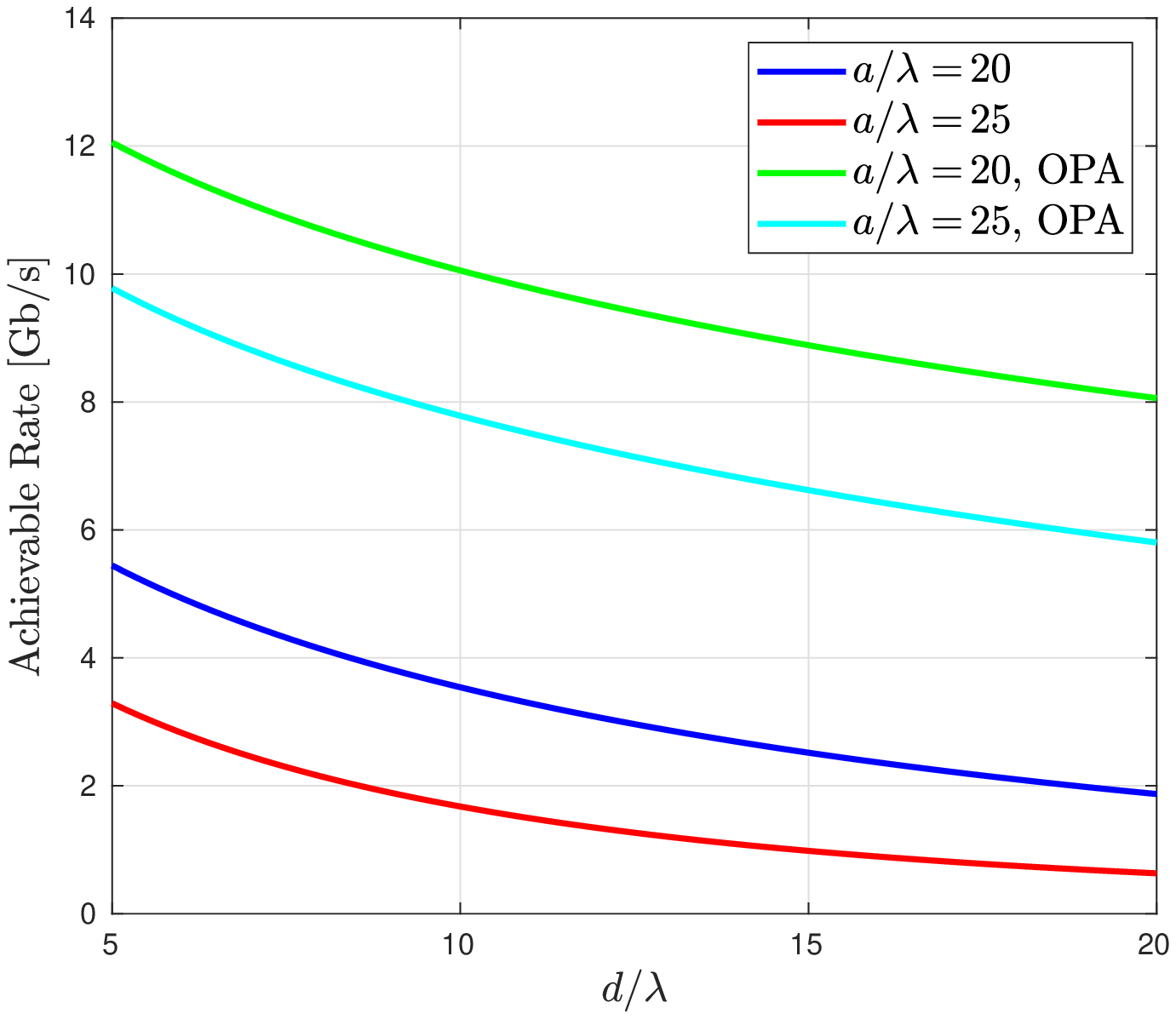} 
    \label{fig:achievable-rate-FF-PA}}}
    \caption{Plots of the achievable rate with and without optimal power allocation (OPA) as a function of $d/\lambda$ for the parallel and colinear configurations with $f_c = 25\, [\text{GHz}]$, $W = 0.2\,f_c$ and $P_{\textrm{max}} = 10\, [\text{mW}]$ in the near- and far-field regions. The far-field plot in (b) corresponds to both parallel and colinear configurations.}
    \label{fig:achievable-rate-PA} 
\end{figure}

Here, we study the impact of the $\textrm{SNR}$, or equivalently the transmit power $P_{\textrm{t}}(f)$, on the performance of compact antennas under optimal power control. Fig. \ref{fig:achievable-rate-PA} depicts the achievable rate for two antenna sizes, $a$, with $\lambda/a\in\{20,25\}$. The horizontal axis shows the ratio $d/\lambda$ for both uniform power allocation, i.e., $P_{\textrm{t}}(f) = P_{\textrm{max}}/W$, as well as optimal power allocation (OPA) $P_{\textrm{t}}(f) =P_{\textrm{t}}^{\tiny{\starletfill}}(f)$ given in (\ref{eq:power-allocation-optimization-solution}). The bandwidth is fixed to $W=0.2\,f_c$ with $f_c = 25\,[\textrm{GHz}]$ and the maximum transmit power is $P_{\textrm{max}} = 10\,[\textrm{mW}]$. The usual crossing between the colinear and parallel configurations is again observed at $d\approx 0.38\,\lambda$ which separates the near- and far-field regions. It can be noticed that the optimal power allocation provides a gain of up to $50\%$ in terms of achievable rate in the near-field region and over $200\%$ in the far-field region.

\section{Conclusion}
In this paper, we analyzed a SISO communication system using the achievable rate criterion by jointly accounting for the physical size limitation of the transmit and receive antennas, and the mutual coupling effects. By employing a circuit-theoretic approach, we incorporated the effects of mutual coupling and the finite-size of Chu's CMS transmit/receive antennas in the channel response and the noise PSD. After establishing an equivalence between Chu antennas and Hertz dipoles, we computed both the near- and far-field mutual impedances using the induced EMF method and the Friis' equation, respectively. This led to a full characterization of the input/output relationship of the proposed circuit-equivalent model. We also examined the effect of the mutual coupling on the achievable rate for both colinear and parallel relative orientations of the transmit and receive antennas under both uniform and optimal power allocation strategies. The analysis presented in this paper can be further extended to colinear/parallel MIMO systems using multi-port circuit theory where the benefit of spatial multiplexing can be investigated under the mutual coupling effect. For future work, it would be interesting to go beyond the assumption that the antennas are both linear time invariant (i.e., without active components, switches, and moving parts) and linearly polarized.


\begin{appendices}
\renewcommand{\thesectiondis}[2]{\roman{section}:}
\section{The Chu-Hertz radiation resistance equivalence}\label{appendix:Chu-Hertz-equivalence}
In this appendix, we show the equivalence between the radiation of the Chu's electric antenna $\Re\big[Z^{\textrm{Chu}}\big]$ and the Hertz dipole $\Re\big[Z^{\textrm{Hertz}}\big]$, which allows us to use them interchangeably as needed throughout the paper. To this end, we perform the following three steps:
\begin{enumerate}[leftmargin=*]
    \item Write the radiated EM fields of both antennas in spherical coordinates ($r$, $\theta$, $\phi$),
    \item Apply the equivalence principle on the equivalent current distributions $\mathbf{J}_{\textrm{Chu}}$ and $\mathbf{J}_{\textrm{Hertz}}$ impressed by a Chu's electric antenna and a Hertz dipole on an enclosed sphere $S$. In doing so, it will be possible to characterize the radiated EM fields of the Chu's antenna by finding the value of the complex coefficient $A_1$ of the $\textrm{TM}_1$ mode,
    \item Use the value of $A_1$ to recover the equality between $\Re\big[Z^{\textrm{Chu}}\big]$ and $\Re\big[Z^{\textrm{Hertz}}\big]$ by computing the radiated power.
\end{enumerate}
\subsubsection{The radiated EM fields of the Chu antenna}\label{appendix:Chu-radiated-EM-fields}
For an arbitrary current distribution enclosed in a sphere of radius $a$, Chu has shown in \cite{chu1948physical} that its radiated EM fields are expressed in terms of a complete set of spherical waves (each of which corresponding to a $\text{TM}_\text{n}$ mode) as follows:
\begin{subequations}\label{eq:Chu-EM-field}
    \begin{align}
    H_{\phi}&~=~\sum_{n=1}^{\infty} A_n \,P_n^1(\cos \theta)\, h_n(kr),\\
E_{r}&~=~-j \,\sqrt{\frac{\mu}{\epsilon}}\,\sum_{n=1}^{\infty} A_n \,n(n+1)\,P_n(\cos \theta) \,\frac{h_n(kr)}{kr},\\
E_{\theta}&~=~j \,\sqrt{\frac{\mu}{\epsilon}}\,\sum_{n=1}^{\infty} A_n \,P_n^1(\cos \theta) \,\frac{1}{kr}\,\frac{\partial}{\partial r}\big(r\, h_n(kr)\big).
    \end{align}
\end{subequations}
In (\ref{eq:Chu-EM-field}), $P_n(\cos \theta)$ and $P_n^1(\cos \theta)$ refer to the Legendre polynomial of order $n$ and the associated Legendre polynomial of first kind, respectively. In addition, $h_n(kr)$ denotes the spherical Hankel function of the second kind, and $A_n$ represents the complex coefficient of the $\text{TM}_\text{n}$ mode.

\noindent For the Chu's electric antenna operating at the $\text{TM}_1$ mode (i.e., $n=1$), the expressions of the EM fields (\ref{eq:Chu-EM-field}) are explicitly given by:
\begin{subequations}\label{eq:electric-Chu-EM-field}
    \begin{align}
    H_{\phi}&~=~ \frac{A_1}{k} \,\sin\theta\,~ \frac{e^{-jkr}}{r}\,\bigg(1+\frac{1}{jkr}\bigg),\label{eq:Chu-H-phi}\\
E_{\theta}&~=~\sqrt{\frac{\mu}{\epsilon}}\, \frac{A_1}{k}\,\sin\theta~ \, \frac{e^{-jkr}}{r}\,\bigg(1 + \frac{1}{j kr} - \frac{1}{(kr)^2}\bigg),\\
E_{r}&~=~2j\,\sqrt{\frac{\mu}{\epsilon}}\, \frac{A_1}{k}\,\cos\theta~ \,\frac{e^{-jkr}}{r}\,\bigg(\frac{1}{kr}+\frac{1}{j(kr)^2}\bigg).
    \end{align}
\end{subequations}

\subsubsection{The radiated EM fields of the Hertz dipole antenna}

\begin{subequations}\label{eq:Hertz-dipole-EM-field}
The radiated EM fields of a Hertz dipole having a uniform current distribution, $I$, over its infinitesimal length, $\text{d}l$, are given by \cite{balanis}:
\begin{align}
    H_{\phi}&~=~jk \,\frac{I \text{d}\ell}{4 \pi}\,\sin\theta\,~ \frac{e^{-j k r}}{r} \,\left(1+\frac{1}{j k r}\right),\label{eq:Hertz-H-phi}\\
    E_{\theta}&~=~jk\eta\,\frac{I \text{d}\ell}{4 \pi} \,\sin\theta\,~\frac{e^{-j k r}}{r}\,\left(1+\frac{1}{j k r}-\frac{1}{(k r)^{2}}\right),\\
    E_r&~=~k\eta\,\frac{I \text{d}\ell}{2 \pi}\cos\theta~\,\frac{e^{-j \beta r}}{r} \left(\frac{1}{kr}+ \frac{1}{j(kr)^{2}}\right).
    \end{align}
\end{subequations}

\subsubsection{The relationship between the equivalent current distributions $\mathbf{J}_{\textrm{Chu}}$ and $\mathbf{J}_{\textrm{Hertz}}$}
\begin{figure}[h!]
    \centering
    \subfloat[]{{
    \includegraphics[scale=0.22]{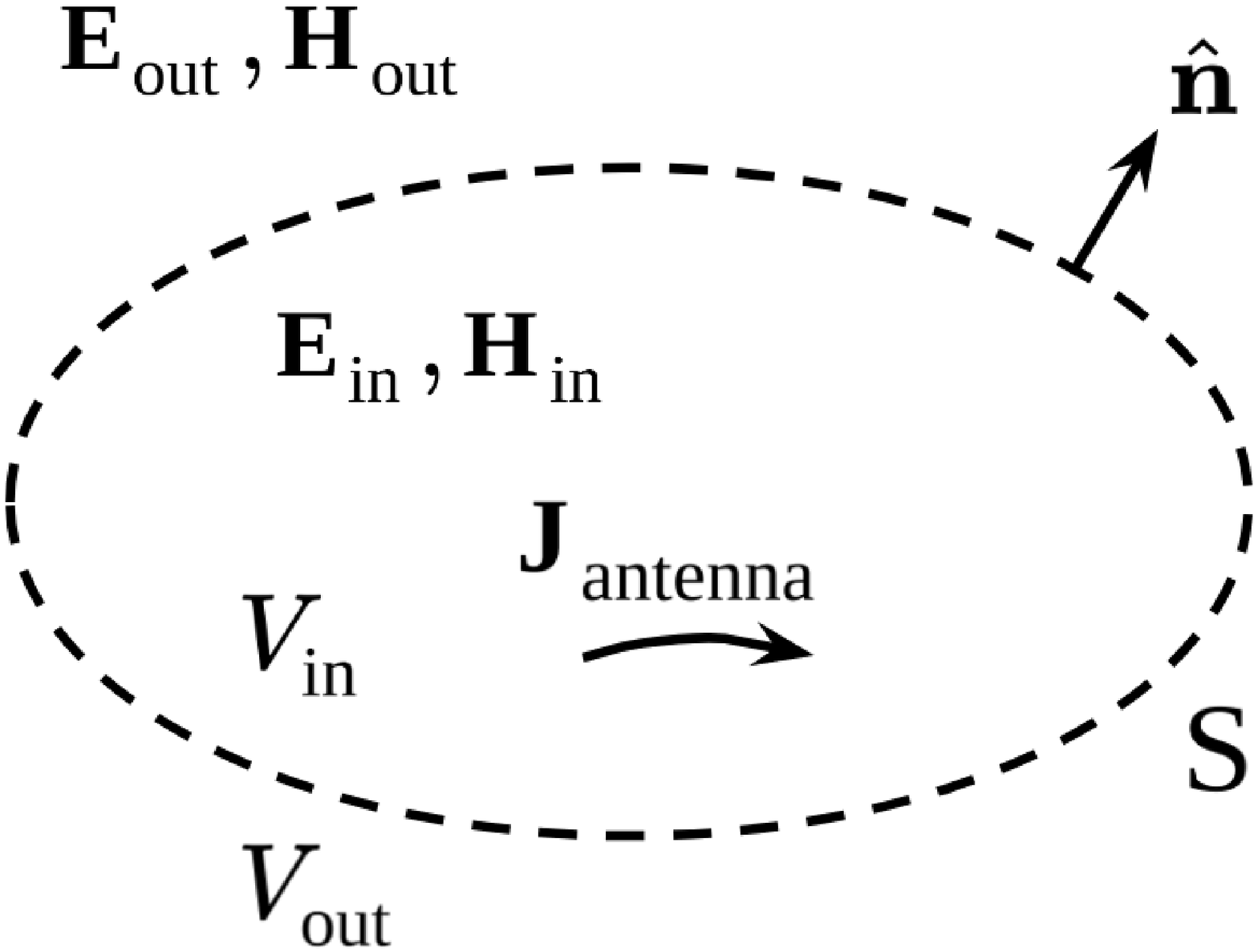} 
    \label{fig:equivalent-theorem-scenario1}}}%
    \qquad~~
    \subfloat[]{{
    \includegraphics[scale=0.22]{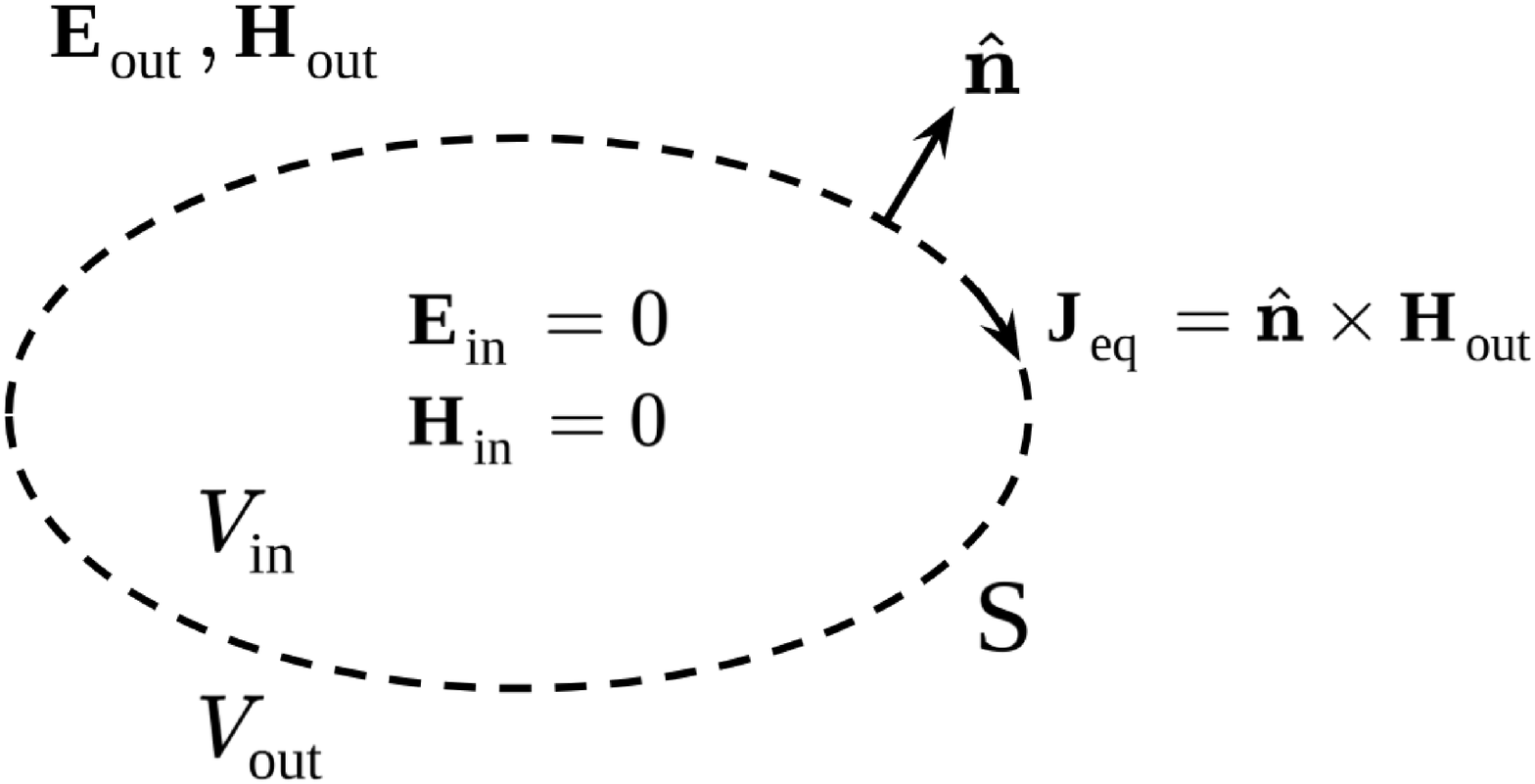}
    \label{fig:equivalent-theorem-scenario1}}}%
    \caption{The equivalence principle substitutes the current distribution $\mathbf{J}_{\textrm{antenna}}$ of an arbitrary antenna in (a) with its equivalent current $\mathbf{J}_{\textrm{eq}} = \hat{\mathbf{n}}\times\mathbf{H}_{\textrm{out}}$ impressed on the surface $S$ in (b) to produce the same fields $\mathbf{E}_{\textrm{out}}$ and $\mathbf{H}_{\textrm{out}}$ outside $\bm{V}_{\textrm{out}}$ when the inside fields $\mathbf{E}_{\textrm{in}}$ and $\mathbf{H}_{\textrm{in}}$ in $\bm{V}_{\textrm{in}}$ are $\mathbf{0}$.}
    \label{fig:equivalence-principle}
\end{figure}

Given the EM fields \big($\mathbf{E}_{\textrm{Chu}}$, $\mathbf{H}_{\textrm{Chu}}$\big) in (\ref{eq:electric-Chu-EM-field}) and \big($\mathbf{E}_{\textrm{Hertz}}$, $\mathbf{H}_{\textrm{Hertz}}$\big)  in (\ref{eq:Hertz-dipole-EM-field}), finding the relationship between the equivalent current distributions $\mathbf{J}_{\textrm{Chu}}$ and $\mathbf{J}_{\textrm{Hertz}}$ boils down to applying the equivalence principle \cite[Chapter 12]{balanis} depicted in Fig. \ref{fig:equivalence-principle}. When the current sources on the structure of the Chu's antenna and the Hertz dipole are replaced by fictitious equivalent sources $\mathbf{J}_{\textrm{eq}} = \hat{\mathbf{n}}\times\mathbf{H}_{\textrm{out}}$, they produce the same EM fields \big($\mathbf{E}_{\textrm{out}}$, $\mathbf{H}_{\textrm{out}}$\big) in the outside volume $V_{\textrm{out}}$.

\noindent Therefore, requiring the equivalent current sources $\mathbf{J}_{\textrm{Chu}}$ and $\mathbf{J}_{\textrm{Hertz}}$ to be identical yields:
\begin{equation}
\hat{\mathbf{n}}\times\mathbf{H}_{\textrm{Chu}} = \hat{\mathbf{n}}\times\mathbf{H}_{\textrm{Hertz}},
\end{equation}
which, after equating (\ref{eq:Chu-H-phi}) and (\ref{eq:Hertz-H-phi}), leads to the condition:
\begin{equation}\label{eq:condition-A1}
    A_1 = jk^2\,\frac{I \text{d}\ell}{4\pi}.
\end{equation}
Now, we recall the definition of the radiation resistance $\Re\big[Z^{\textrm{Hertz}}\big]$ of the Hertz dipole when regarded as an impedance $Z^{\textrm{Hertz}}$:
\begin{equation}\label{eq:Hertz-radiation-resistance}
    \Re\big[Z^{\textrm{Hertz}}\big] =\frac{2\pi}{3} \,\eta\,\frac{\textrm{d}\ell^2}{\lambda^2}.
\end{equation}
Upon rearrangement of (\ref{eq:Hertz-radiation-resistance}) and substitution for $\textrm{d}\ell$ in (\ref{eq:condition-A1}), the final expression of the complex coefficient $A_1$ becomes:
\begin{equation}\label{eq:Hertz-Z-A1}
    A_1 = j\,\frac{I k^2 c}{4\pi f} \sqrt{\frac{3\, \Re\big[Z^{\textrm{Hertz}}\big]}{2\pi\eta}}.
\end{equation}
Now, given the $\textrm{TM}_1$ mode coefficient $A_1$ in (\ref{eq:Hertz-Z-A1}), the radiated EM fields \big($\mathbf{E}_{\textrm{Chu}}$, $\mathbf{H}_{\textrm{Chu}}$\big) are fully characterized according to (\ref{eq:electric-Chu-EM-field}).
\subsubsection{The equality between the  radiation  of  the  electric  Chu’s  antenna $\Re\big[Z^{\textrm{Chu}}\big]$ and the  Hertz  dipole $\Re\big[Z^{\textrm{Hertz}}\big]$}
To compute the radiation resistance $\Re\big[Z^{\textrm{Chu}}\big]$ of the Chu's electric antenna as a function of the radiation resistance $\Re\big[Z^{\textrm{Hertz}}\big]$ of the Hertz dipole, we will first calculate the total radiated power $P_{\textrm{Chu}}$ by the Chu's electric antenna on a sphere of radius $a$. This can be achieved by integrating the energy density $\mathbf{S} =\frac{1}{2}\, \Re\big[\mathbf{E}\times\xbar{\mathbf{H}}\,\big]$ over the solid angle of a sphere of radius $a$ as follows:
\begin{equation}\label{eq:power-chu-antenna}
\begin{aligned}[b]
P_{\textrm{Chu}}(a)=\int_{\text {sphere }} \mathbf{S} \,\cdot\, \text{d} \mathbf{S}&=\frac{1}{2}\int_{0}^{2 \pi} \int_{0}^{\pi} \Big(E_{\theta} \,\xbar{H_{\phi}} -E_{r} \,\xbar{H_{\phi}}\Big) \, a^2\,\sin\theta\, \textrm{d}\theta \,\textrm{d} \phi = \frac{4\pi}{3}\,\frac{\eta}{k^2}\,|A_1|^2,
\end{aligned}
\end{equation}
where $\textrm{d}\mathbf{S} = \widehat{\bm{r}}\, \sin(\theta)\,\textrm{d}\theta\,\textrm{d}\phi$ is the normal surface vector. We then identify the radiation resistance $\Re\big[Z^{\textrm{Chu}}\big]$ by treating the Chu's antenna as a resistance with a power loss $P_{\textrm{Chu}} = \frac{1}{2}\,I^2\,\Re\big[Z^{\textrm{Chu}}\big]$, which is equated with (\ref{eq:power-chu-antenna}) to yield the following relationship between $\Re\big[Z^{\textrm{Chu}}\big]$ and $A_1$:
\begin{equation}\label{eq:relationship-Z-chu-A1}
|A_1|^2 = \frac{3k^2I^2}{8\pi\eta}\, \Re\big[Z^{\textrm{Chu}}\big].
\end{equation}
Finally, we combine (\ref{eq:Hertz-Z-A1})  and (\ref{eq:relationship-Z-chu-A1}) to obtain $\Re\big[Z^{\textrm{Chu}}\big] = \Re\big[Z^{\textrm{Hertz}}\big]$, thereby confirming the identical radiated EM fields for both the Hertz dipole and the electrical Chu's antenna.
\section{The Chu-Hertz mutual impedance equivalence}\label{appendix:self-mutula-impedance-SISO}
\subsection{Mutual impedance between two Hertz dipoles}\label{appendix:self-mutual-impedance-SISO-Hertz}
In this appendix, we derive the mutual impedances between two electrical Chu's antennas in close proximity which characterise the off-diagonal entries of the impedance matrix describing a SISO near-field communication channel.

\noindent Given a current $I_0$ that is uniformly distributed along its length $\text{d}l$, a Hertz dipole radiates an electric field in spherical coordinates ($r$, $\theta$, $\phi$) which is given by:
\begin{subequations}\label{eq:hertz-dipole-E-field}
    \begin{align} 
        E_{\theta} &= j\,\frac{\eta_0\,k_0\,I_0\,\text{d}\ell}{4\pi}\, \Bigg(1+\frac{1}{j\,k_0\,r} - \frac{1}{(k_0\,r)^2}\Bigg)\,\frac{e^{-j\,k_0\,r}}{r}\,\sin(\theta),\\
        E_r &= \frac{\eta_0\,I_0\,\text{d}\ell}{2\pi}\, \Bigg(\frac{1}{r}-\frac{j}{k_0\,r^2}\Bigg)\,\frac{e^{-j\,k_0\,r}}{r}\,\cos(\theta),\\
        E_{\phi} &= 0.
    \end{align} 
\end{subequations}
Let $\text{Hertz}_\text{T}$ and $\text{Hertz}_\text{R}$ denote two transmit and receive Hertz dipole antennas of length $\text{d}l_\text{T}$ and $\text{d}l_\text{R}$ and radiation resistance $\Re\big[Z^{\textrm{Hertz}_\text{T}}\big]$ and $\Re\big[Z^{\textrm{Hertz}_\text{R}}\big]$, respectively, which are located at an arbitrarily fixed plane and separated by a distance $d$ as shown in Fig. \ref{fig:two-hertz-antennas}. $\text{Hertz}_\text{R}$ is oriented along a chosen w-axis whose direction vector can be written as $\uvec w = \cos(\gamma) \,\uvec r \,+\, \sin(\gamma)\,\boldsymbol{\hat{{\theta}}}$. The orientation of $\text{Hertz}_\text{T}$ coincides with the z-axis to simplify the calculation of the electric field. Such configuration corresponds to $\text{Hertz}_\text{T}$ and $\text{Hertz}_\text{R}$ being aligned w.r.t. their axes $w$ and $z$, respectively, and arbitrarily rotated with angles $\beta$ and $\gamma$ w.r.t. their connecting axis $r$.


\noindent To compute the mutual impedance $Z^{\text{Hertz}}_{\text{TR}}$, one should consider the observation point over $\text{Hertz}_\text{R}$ along $\uvec w$. When $\text{Hertz}_\text{R}$ is open-circuited, the relationship between the induced open-circuit voltage of $Z_{\text{TR}}$, denoted as $V_{\textrm{TR,oc}}$, and the electric field $E_{\text{TR}}(r,\theta)$ generated by $\text{Hertz}_\text{T}$ and incident on $\text{Hertz}_\text{R}$ is
\begin{equation}\label{eq:open-circuit-voltage-A2}
    \begin{aligned}[b]
        V_{\textrm{TR,oc}} &= -\frac{1}{I_{0,\text{R}}} \int\hspace{-0.32cm}\int_{\text{Hertz}_\text{R}}E_{\text{TR}}(r,\theta) \, I_{0,\text{R}}(r,\theta)\,\text{d}r\,\text{d}\theta\\
        &= - \text{d}l_\text{R}\Big[E_{r}(r,\theta)\,\cos(\gamma)+ E_{\theta}(r,\theta)\,\sin(\gamma)\Big]\biggr\rvert_{r= d, \,\theta=\beta}\\
        &= -\frac{\eta_0\,I_{0,\text{T}}\,\text{d}\ell_\text{T}\,\text{d}\ell_\text{R}\,k_0^2}{2\pi} \Bigg[\frac{1}{2}\,\sin(\beta)\,\sin(\gamma)\bigg(\frac{1}{j\,k_0\,d} + \frac{1}{(j\,k_0\,d)^2} + \frac{1}{(j\,k_0\,d)^3}\bigg)\\
        & \hspace{4cm}+ \cos(\gamma)\,\cos(\beta)\bigg(\frac{1}{(j\,k_0\,d)^2} + \frac{1}{(j\,k_0\,d)^3}\bigg)\Bigg]\,e^{-j\,k_0\,d},
    \end{aligned}
\end{equation}
where we carried out the integration explicitly because the length $\text{d}l_{\text{R}}$ of the Hertz dipole is infinitesimal. Using (\ref{eq:open-circuit-voltage-A2}), the mutual impedance $Z^{\text{Hertz}}_{\text{RT}}$ is 
\begin{equation}\label{eq:impedance-Z-21}
    \begin{aligned}[b]
        Z^{\text{Hertz}}_{\text{RT}} &=\frac{V_{\textrm{TR,oc}}}{I_{0,\text{T}}}\\
        &= -\frac{\eta_0\,\text{d}\ell_\text{T}\,\text{d}\ell_\text{R}\,k_0^2}{2\pi} \Bigg[\frac{1}{2}\,\sin(\beta)\,\sin(\gamma)\bigg(\frac{1}{j\,k_0\,d} + \frac{1}{(j\,k_0\,d)^2} + \frac{1}{(j\,k_0\,d)^3}\bigg)\\
        & \hspace{4cm}+ \cos(\beta)\,\cos(\gamma)\bigg(\frac{1}{(j\,k_0\,d)^2} + \frac{1}{(j\,k_0\,d)^3}\bigg)\Bigg]\,e^{-j\,k_0\,d}.
    \end{aligned}
\end{equation}
Substituting for $\text{d}l_\text{R}$ and $\text{d}l_\text{T}$ in (\ref{eq:impedance-Z-21}) using (\ref{eq:Hertz-radiation-resistance}) yields the final expression
\begin{equation}\label{eq:impedance-Z-21-final}
    \begin{aligned}[b]
        Z_{\text{RT}}^{\text{Hertz}} &=-\frac{3\,k_0^2\,c^2}{4\pi^2f^2}\,\sqrt{\Re\big[\mathbf{Z}^{\textrm{Hertz}_\text{T}}\big]\,\Re\big[\mathbf{Z}^{\textrm{Hertz}_\text{R}}\big]} \Bigg[\frac{1}{2}\,\sin(\beta)\,\sin(\gamma)\bigg(\frac{1}{j\,k_0\,d} + \frac{1}{(j\,k_0\,d)^2} + \frac{1}{(j\,k_0\,d)^3}\bigg)\\
        & \hspace{4cm}+ \cos(\beta)\,\cos(\gamma)\bigg(\frac{1}{(j\,k_0\,d)^2} + \frac{1}{(j\,k_0\,d)^3}\bigg)\Bigg]\,e^{-j\,k_0\,d}.
    \end{aligned}
\end{equation}
Note that $Z_{\text{TR}} = Z_{\text{RT}}$, due to the reciprocity theorem when applied on $\text{Hertz}_\text{T}$ and $\text{Hertz}_\text{R}$.

\subsection{Mutual impedances of Chu's antennas}\label{appendix:self-mutual-impedance-Chu}
Given the mutual impedance between two Hertz dipoles (\ref{eq:impedance-Z-21-final}) and the equivalence between Chu's electrical antennas and Hertz dipoles established in Appendix \ref{appendix:Chu-Hertz-equivalence} in terms of radiation resistance, it is possible to deduce the mutual impedance between two electrical Chu's antennas without recalculating them from scratch.

\noindent To this end, we consider two near-field SISO communication scenarios:
\begin{itemize}
    \item \textit{Scenario 1:} an electrical Chu's antenna as a transmitter to a receiving Hertz dipole antenna,
    \item \textit{Scenario 2:} a Hertz dipole antenna as a transmitter to a receiving electrical Chu's antenna.
\end{itemize}
For the first scenario, we establish a relationship between the current of the electrical Chu's antenna, $I^{\textrm{Chu}}_{\text{T}}$, and the current of the Hertz dipole, $I^{\textrm{Hertz}}_{\text{R}}$. The current of the electrical Chu's antenna is given by \cite{chu1948physical}:
\begin{equation}\label{eq:I-chu}
\begin{aligned}
 I^{\textrm{Chu}}_{\text{T}} &= - \sqrt{\frac{8\pi\eta}{3}}\,\frac{A_1}{k}  \,\frac{ka-j}{ka}\,e^{-jka}.
\end{aligned}
\end{equation}
After injecting the expression of $A_1$ established in (\ref{eq:condition-A1}) into (\ref{eq:I-chu}) and computing its squared norm, one obtains:
\begin{equation}\label{eq:I-chu-squared}
\begin{aligned}
 (I_{\text{T}}^{\textrm{Chu}})^2 &= \frac{\eta \,k^2\,d\ell_{\text{T}}^2}{6\pi}\, (I_{\text{R}}^{\textrm{Hertz}})^2\,\frac{(ka)^2 +1}{(ka)^2}.
\end{aligned}
\end{equation}
Using the definition of the self-impedance of the Chu's electric antenna (\ref{eq:Z1-chu}) and substituting $\text{d}l_{\text{T}}$ with its expression from (\ref{eq:Hertz-radiation-resistance}), (\ref{eq:I-chu-squared}) reduces simply to:
\begin{equation}
    \Bigg(\frac{I_{\text{T}}^{\textrm{Chu}}}{I_{\text{R}}^{\textrm{Hertz}}}\Bigg)^2 = \frac{\Re\big[Z^{\textrm{Hertz}}_{\text{R}}\big]}{\Re\big[Z^{\textrm{Chu}}_{\text{T}}\big]},
\end{equation}
or equivalently:
\begin{equation}\label{eq:I-chu-I-hertz}
     I_{\text{T}}^{\textrm{Chu}} = \sqrt{\frac{\Re\big[Z^{\textrm{Hertz}}_{\text{R}}]}{\Re\big[Z^{\textrm{Chu}}_{\text{T}}]}}\,I^{\textrm{Hertz}}_\text{T}.
\end{equation}
Similarly, if we consider the second scenario (i.e., by swapping the types of the transmit and receive antennas in the first scenario), we obtain in pure analogy to (\ref{eq:I-chu-I-hertz}) the following relationship:
\begin{equation}\label{eq:I-chu-I-hertz-2}
     I^{\textrm{Hertz}}_{\text{T}} = \sqrt{\frac{\Re\big[Z^{\textrm{Chu}}_{\text{R}}\big]}{\Re\big[Z^{\textrm{Hertz}}_{\text{T}}\big]}}\,I_{\text{T}}^{\textrm{Chu}},
\end{equation}
which, when combined with (\ref{eq:I-chu-I-hertz}) yields:
\begin{equation}\label{eq:I-chu-I-hertz-final}
(I_{\text{T}}^{\textrm{Chu}})^2 \, \sqrt{\Re\big[Z^{\textrm{Chu}}_{\text{T}}\big] \, \Re\big[Z^{\textrm{Chu}}_{\text{R}}\big]} = (I_{\text{T}}^{\textrm{Hertz}})^2 \, \sqrt{\Re\big[Z^{\textrm{Hertz}}_{\text{T}}\big] \, \Re\big[Z^{\textrm{Hertz}}_{\text{R}}\big]}.
\end{equation}
The identity in (\ref{eq:I-chu-I-hertz-final}) can now be used to deduce the relationship between the mutual impedance of two Hertz dipoles, $Z^{\text{Hertz}}_{\textrm{TR}}$, and the mutual impedance of two electrical Chu's antennas, $Z^{\text{Chu}}_{\textrm{TR}}$.

\noindent In fact, for two Hertz dipole antennas, the mutual impedance between the first and second antennas is given by:
\begin{equation}\label{eq:mutual-impedance-hertz-definition}
    Z^{\textrm{Hertz}}_{\text{TR}} = -\frac{1}{(I^{\textrm{Hertz}})^2} \int E_{\text{TR}}^{\textrm{Hertz}}(r,\theta) \, I^{\textrm{Hertz}}(r,\theta)\,\text{d}r\,\text{d}\theta.
\end{equation}
Since the electrical Chu's antenna and the Hertz dipole have the same radiated EM fields, as well as, the same current distribution $\mathbf{J}_{\textrm{Chu}}$ and $\mathbf{J}_{\textrm{Hertz}}$ as already shown in Appendix~\ref{appendix:Chu-Hertz-equivalence},  one can replace $E^{\textrm{Hertz}}_{\text{TR}}(r,\theta)$ with $E_{\text{TR}}^{\textrm{Chu}}(r,\theta)$, and $I^{\textrm{Hertz}}(l)$ with $I^{\textrm{Chu}}(l)$. By doing so, (\ref{eq:mutual-impedance-hertz-definition}) leads to:
\begin{equation}\label{eq:mutual-impedance-chu-definition}
    \mathbf{Z}^{\textrm{Hertz}}_{\text{TR}} = -\frac{1}{(I^{\textrm{Hertz}})^2} \int E_{\text{TR}}^{\textrm{Chu}}(r,\theta) \, I^{\textrm{Chu}}(r,\theta)\,\text{d}r\,\text{d}\theta,
\end{equation}
which yields, after normalizing both sides with the same factor:
\begin{equation}\label{eq:mutual-impedance-derivation}
\begin{aligned}
\frac{Z^{\textrm{Hertz}}_{\text{RT}}}{\sqrt{\Re\big[Z^{\textrm{Hertz}}_{\text{T}}\big] \, \Re\big[Z^{\textrm{Hertz}}_{\text{R}}\big]}} &=-\frac{1}{(I^{\textrm{Hertz})^2}\,\sqrt{\Re\big[Z^{\textrm{Hertz}}_{\text{T}}\big] \, \Re\big[Z^{\textrm{Hertz}}_{\text{R}}\big]}} \int E_{\text{TR}}^{\textrm{Chu}}(r,\theta) \, I_{\textrm{Chu}}(r,\theta)\,\text{d}r\,\text{d}\theta.
\end{aligned}
\end{equation}

\noindent After substituting the denominator of the right-hand side of (\ref{eq:mutual-impedance-derivation}) with (\ref{eq:I-chu-I-hertz}), it follows that:
\begin{equation}\label{eq:mutual-impedance-derivation-final}
\begin{aligned}
\frac{Z^{\textrm{Hertz}}_{\text{RT}}}{\sqrt{\Re\big[Z^{\textrm{Hertz}}_{\text{T}}\big] \, \Re\big[Z^{\textrm{Hertz}}_{\text{R}}\big]}} &=\frac{Z^{\textrm{Chu}}_{\text{RT}}}{\sqrt{\Re\big[Z^{\textrm{Chu}}_{\text{T}}\big] \, \Re\big[Z^{\textrm{Chu}}_{\text{R}}\big]}},
\end{aligned}
\end{equation}
which allows one to deduce the mutual impedance $Z^{\textrm{Chu}}_{\text{RT}}$ between two electrical Chu's antennas given the mutual impedance $Z^{\textrm{Hertz}}_{\text{RT}}$ and the self-impedances of the two antennas. Owing to the reciprocity theorem, by switching the roles of the transmit and receive antennas, one can establish the equality in (\ref{eq:mutual-impedance-derivation-final}) using $Z^{\textrm{Hertz}}_{\text{TR}}$ and $Z^{\textrm{Chu}}_{\text{TR}}$ instead of $Z^{\textrm{Hertz}}_{\text{RT}}$ and $Z^{\textrm{Chu}}_{\text{RT}}$, thereby obtaining:
\begin{equation}\label{eq:mutual-impedance-derivation-final2}
\begin{aligned}
\frac{Z^{\textrm{Hertz}}_{\text{TR}}}{\sqrt{\Re\big[Z^{\textrm{Hertz}}_{\text{T}}\big] \, \Re\big[Z^{\textrm{Hertz}}_{\text{R}}\big]}} &=\frac{Z^{\textrm{Chu}}_{\text{TR}}}{\sqrt{\Re\big[Z^{\textrm{Chu}}_{\text{T}}\big] \, \Re\big[Z^{\textrm{Chu}}_{\text{R}}\big]}}.
\end{aligned}
\end{equation}
\section{The derivation of the far-field mutual impedance}\label{appendix:FF-mutual-impedance}
We resort to basic circuit theory analysis of the FF SISO communication model described in Fig.~\ref{fig:siso-far-field-system-model-nomn} to find the FF mutual impedance, $Z_{\mathrm{RT}}^{\textrm{Chu}}(f)$. Using the current divider, $I_{R_1}(f)$ and $I_1(f)$ are related as follows:
\begin{equation}\label{eq:relations-current-dividers}
    I_{R_1}(f)=I_1(f)\,\left( \frac{j\,2\pi\,f\,a_{\textrm{T}}}{c + j\,2\pi\,f\,a_{\textrm{T}}} \right)\,[\text{A}].
\end{equation}
To find the transmit-receive $Z_{\mathrm{RT}}^{\textrm{Chu}}(f)$ that relates the receiver voltage $V_2(f)$ and the transmitter current $I_1(f)$, we first open-circuit the receiver current by setting $I_2(f) = 0$. The equivalent impedance $Z_{R_2\,\parallel\,L}$ of the parallel-connected resistor $R_2$ and the inductance $L = a_{\textrm{R}}\,R_2/c$ is then given by:
\begin{equation}
    Z_{R_2\,\parallel\,L}(f) = \frac{j\,2\pi\,f\,R_2}{j\,2\pi\,f + \frac{c}{a_{\textrm{R}}}}\,[\Omega].
\end{equation}
By recalling the expression of $|I_s(f)|$ in (\ref{eq:friss-current-magnitude}) and using Ohm's law for the impedance $Z_{R_2\,\parallel\,L}(f)$, i.e., $V_2(f) = Z_{R_2\,\parallel\,L}(f) \, I_s(f)$ one obtains:
\begin{equation}\label{eq:V2-formula}
    V_2(f) = I_{R_1}(f)\,\frac{c}{2{\pi}fd}\,\sqrt{\frac{G_\textrm{T}\,G_\textrm{R}\,R_1}{R_2}}\,\frac{j\,2\pi\,f\,R_2}{j\,2\pi\,f + \frac{c}{a_{\textrm{R}}}}\,[\text{V}].
\end{equation}
Injecting (\ref{eq:relations-current-dividers}) into (\ref{eq:V2-formula}) yields:
\begin{equation}
    V_2(f) = I_1(f)\,\left( \frac{j\,2\pi\,f\,a_{\textrm{T}}}{c + j\,2\pi\,f\,a_{\textrm{T}}} \right)\,\frac{c}{2{\pi}fd}\,\sqrt{\frac{G_\text{T}\,G_\text{R}\,R_1}{R_2}}\,\frac{j\,2\pi\,f\,R_2}{j\,2\pi\,f + \frac{c}{a_{\textrm{R}}}}\,[\text{V}],
\end{equation}

\noindent from which we obtain the expression of $Z_{\mathrm{RT}}^{\textrm{Chu}}(f)$ given in (\ref{eq:FF-mutual-impedance}).
\end{appendices}

\bibliographystyle{IEEEtran}
\bibliography{IEEEabrv,references}

\end{document}